# Assessing Power System Vulnerability to Climate-Related Stressors and Shocks: The Case of Indonesia

Hariadi Aji [a, b, *], Nihit Goyal [a], Stefan Pfenninger-Lee [a], Igor Nikolic [a]

[a] Faculty of Technology, Policy and Management, Delft University of Technology, Delft, The Netherlands.

[b] PT PLN (Persero), Jl. Trunojoyo Blok MI / 135, Jakarta Selatan, Indonesia

*** Corresponding author**:

E-mail address: h.aji@tudelft.nl

**Abstract**

Climate change and extreme weather are increasing the vulnerability of power systems globally, particularly in emerging economies such as Indonesia. Yet, existing studies often assess these impacts in isolation, focusing on individual components or specific hazards, leaving system-level implications under-examined. To address this gap, we develop an integrated, spatially explicit approach to assess an energy system's climate-related vulnerabilities and their impacts, and apply the approach to Indonesia. We quantify climate-based vulnerabilities in generation and transmission infrastructure as well as in demand by distinguishing stressors (temperature rise) and shocks (disruptive hazards due to sea-level rise, flooding, and landslides). Through geospatial data analysis, derating models, and regression analysis, we examine existing and planned assets and demand under historical and future climate conditions. Results indicate that both existing and planned generation are likely to experience stress, which implies a reduction in usable capacity, even as electricity demand increases due to temperature rise, and transmission assets face potential disruption from climate change-induced shocks. Together, these effects erode reserve margins by up to 36 percentage points under the 10-year plan, indicating a substantial reduction in system resilience. The largest system, Jawa-Madura-Bali, experiences a 20.8 percentage point decline, leaving a remaining margin of 26.5%, below the 10-year planning threshold. Importantly, the findings suggest that a growing share of future capacity expansion may be absorbed by climate-induced losses, implying that adaptation-related investments may increasingly be required simply to maintain existing supply levels rather than meet future requirements. We conclude that there is an urgent need to embed climate considerations more explicitly into power sector planning.

**Highlights**

1. Indonesia's power system is exposed to chronic climate stressors and acute shocks.
2. Temperature stressors reduce usable capacity and increase electricity demand.
3. Climate shocks add exposure and further tighten system conditions.
4. Combined impacts reduce reserve margins and system adequacy.
5. Future investments may offset climate losses rather than expand supply

**Abbreviations**

| | |
|---|---|
| 10m wind | 10-m wind speed from ERA5 |
| ARED | Accelerated Renewable Energy Development – Scenario in 10-year plan, RUPTL that extends renewable energy supply. |
| BNPB | Badan Nasional Penanggulangan Bencana/ Indonesia's National Disaster Countermeasures Authority |
| CCGT | Combined-cycle gas turbines |
| CMIP5 | Coupled Model Intercomparison Project-5 |
| GCM | Global Climate Model |
| GIS | Geographic Information System |
| GEDTM | Global Ensemble Digital Terrain Model |
| OCGT | Open-cycle gas turbines |
| OLS | Ordinary Least Squares |
| PLN | Perusahaan Listrik Negara / Indonesia Electric Utility Company |
| PV | Photovoltaic |
| RCM | Regional Climate Model |
| RCP | Representative Concentration Pathways |
| REBase | Renewable Energy Base – Scenario in 10-year plan, RUPTL |
| RUKN | Rencana Umum Ketenagalistrikan Nasional / 2060 long-term electricity plan |
| RUPTL | Rencana Usaha Penyediaan Tenaga Listrik / 10-year Indonesian Electricity Business Plan |
| SLR | Sea Level Rise |
| SSP | Shared Socioeconomics Pathways |
| t2m | 2-m air temperature from ERA5 |
| tasmax | Daily maximum near-surface air temperature |

## 1. Introduction

Global ambient temperature has already risen by 1.1°C above pre-industrial levels (1850–1900) during 2011–2020 and is expected to continue rising (IPCC, 2023). Moreover, over the observational record up to 2018, the global mean sea level increased by up to 3.7 mm per year and is projected to continue rising in the future (DeConto et al., 2021; IPCC, 2023). These changes, coupled with the more extreme weather events they cause, pose an increasing threat to power systems (Cronin et al., 2018; Xu et al., 2024). The three main components —generation, transmission, and load demand —are all vulnerable to the impacts of climate change.

Research on climate change impacts on power systems, which has been done in various countries, has examined a wide range of effects, ranging from gradual performance degradation under changing climate conditions (Chandramowli & Felder, 2014) to disruptions caused by extreme weather events (Mujjuni et al., 2023; Panteli & Mancarella, 2015). In terms of gradual performance degradation, studies have quantified impacts on generation efficiency, cooling-water availability, electricity demand, and transmission capacity across multiple regions, for example, the United States (M. D. Bartos & Chester, 2015; Burillo et al., 2019; Szinai et al., 2024), Europe (Ravestein et al., 2018; Van Vliet et al., 2013), and China (Zhang et al., 2021). On the other hand, studies on disruptions have analyzed impacts on grid infrastructure through fragility curves, failure rates, and cascading outages (Mujjuni et al., 2023; Panteli & Mancarella, 2015). Much of this work examines the vulnerability of transmission and distribution infrastructure under storms, hurricanes, and wind shocks, including applications in Texas (Lian et al., 2023), New York (Haggag et al., 2021), and Greece (Gkika et al., 2024). The research mentioned partially analyzes the impacts of climate change on separate power system components. However, integrated assessments of generation, transmission, and demand are still limited in the literature, with a notable gap in country-specific studies focused on Indonesia.

Indonesia, being the fourth most populous country in the world and an emerging economy, relies on growing electricity demand to further its development. Yet climate change and extreme weather events have already started impacting the electricity system in Indonesia (PLN, 2025; Tempo, 2025), indicating that future development must address system resilience. As a developing country, much of Indonesia's power system infrastructure is currently being built. This presents an opportunity to appropriately consider changing climate and weather impacts in the design of new infrastructure, rather than modify the system afterward. However, the electricity plan in Indonesia (MEMR, 2024a, 2025) is currently focused on climate change mitigation through $CO_2$ emission reduction. While awareness and research on climate change adaptation are increasing, adaptation considerations are not systematically reflected in national power sector planning yet (Setiawan et al., 2025). Strengthening research on climate change impacts on the power system represents an important starting point for advancing its adaptation.

Research on climate change impacts in Indonesia's power sector remains largely component-specific. Most studies focus on individual elements, such as temperature-sensitive electricity demand (Veanti et al., 2022) or performance impacts on specific generation technologies, particularly solar PV (Dewi et al., 2019; Tarigan, 2019). While one study has examined climate-related vulnerabilities across power generation, transmission, and demand (Handayani et al., 2019), it relies primarily on qualitative assessments and does not provide quantitative metrics that can be directly integrated into power system planning models. Subsequent work has incorporated climate considerations within a system-level optimization framework, but focused only on the Jawa–Bali system (Handayani et al., 2020), limiting its ability to reflect nationwide vulnerabilities across Indonesia's diverse power systems.

Despite this work, a clear gap remains: No study has systematically assessed the vulnerability of Indonesia's current and planned power system to climate change at a national level, particularly under rising temperatures. In addition, disruptive hazards such as floods, landslides, and sea level rise may further affect system conditions. To address this gap, this study asks: What are the key vulnerabilities of Indonesia's power generation, transmission, and demand to climate change impacts under current conditions and along Indonesia's pathway toward net-zero emissions? Addressing this question requires considering climate impacts across generation, transmission, and demand within Indonesia's interconnected power system, as well as across spatial scales and planning horizons.

This paper is organized as follows. Section 2 provides an overview of Indonesia's power system context, including the physical and governance aspects for extra information about Indonesia. The latter part of the section explains Indonesia's climate background that affects generation, transmission, and demand. Section 3 describes the research methods. Section 4 presents the quantitative results of the vulnerability assessment for the existing power system and the resilience of the planned future system under climate change. Section 5 discusses the interpretations of the findings, policy implications, and limitations. Finally, Section 6 concludes the study. The supplementary materials provides details of the results that support the main findings delivered in the paper. The code used for quantification is also included in the repository to ensure transparency and reproducibility. However, the original component-level electricity asset datasets are not publicly available.

## 2. Indonesia Power System and Climate Context

This section provides background on Indonesia's power system, including its physical structure, governance, and planning framework, and the relevant climate context. It also introduces the climate sensitivity of key power system components, noting that climate exposure depends not only on physical hazards, but also on how power systems are governed and planned.

### *2.1 Physical and Structural Characteristics of Indonesia's Power System*

Indonesia's growing power sector ranks 12$^{th}$ globally in electricity consumption in 2024, reaching 371.6 TWh and a peak load of 61.3 GW (MEMR, 2023). The total existing generation capacity in Indonesia is around 74 GW (MEMR, 2025). Fossil fuels dominate Indonesia's power generation, with approximately 60% coal and 30% natural gas (MEMR, 2024b). Indonesia's archipelagic geography thus results in separate transmission grids. The greatest demand is in Jawa, the most populous island in the world and home to roughly 160 million of Indonesia's 278 million people. It accounts for more than half of the national electricity demand. Other major islands operate their own distinct power systems with separate generation and transmission structures, as illustrated in Fig. 1.

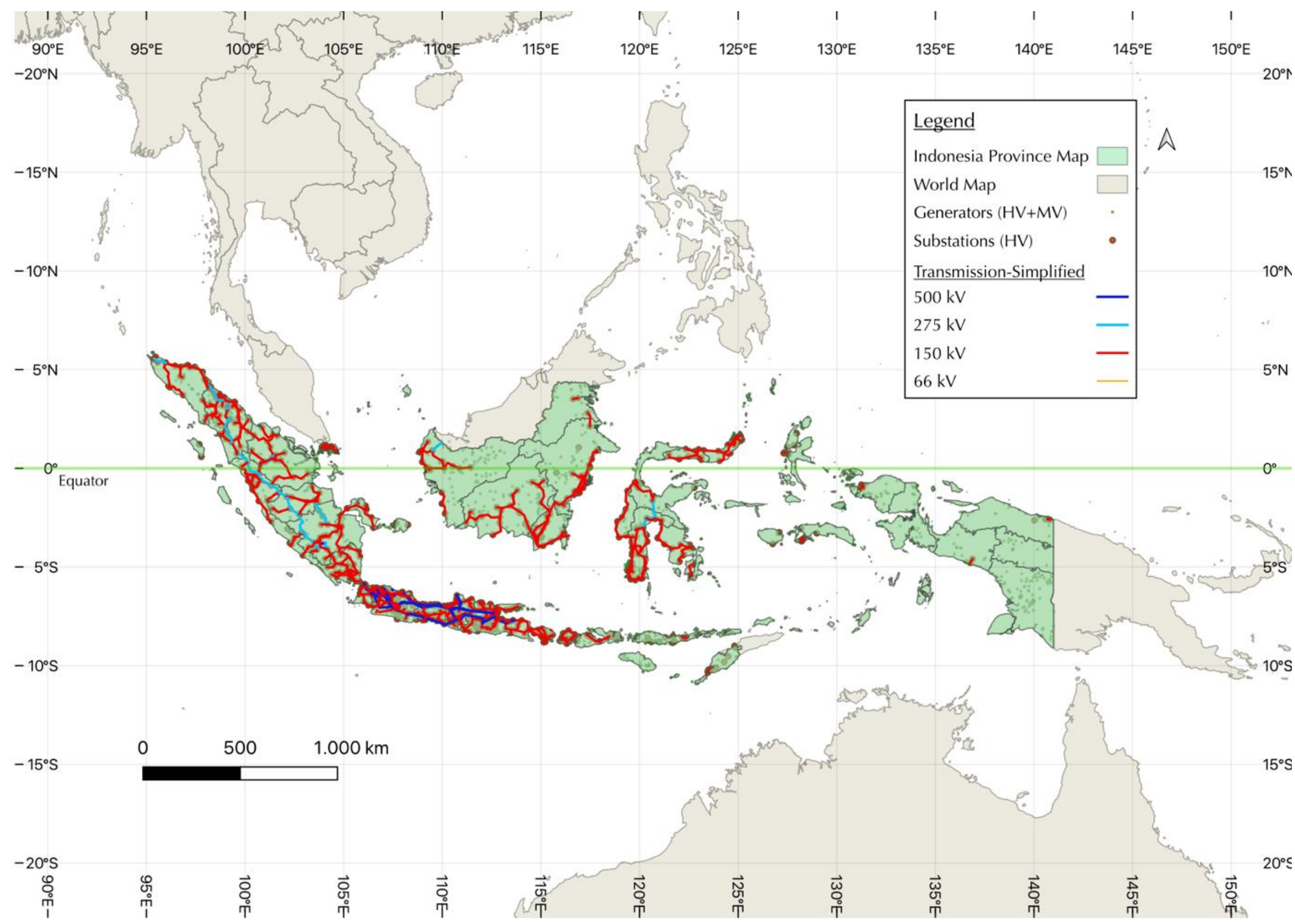

*Fig. 1. Geographical context and power system structure of Indonesia.*

*2.2 Governance and Planning of Indonesia's Power Sector*

Indonesia's power sector operates under a vertically integrated structure, managed by the public utility company, Perusahaan Listrik Negara (PLN), which plans and operates electricity generation, transmission, and distribution. The sector is coordinated by the Ministry of Energy and Mineral Resources (MEMR/ESDM) with recent involvement of the Danantara authority, as part of Indonesia's sovereign wealth fund. The sector operates large assets that predominantly rely on conventional technologies, including centralized, fuel-based generation and transmission infrastructure with limited development of flexible, renewable, and inverter-based resources.

Indonesia's power sector development is guided by two main planning documents: the long-term National Electricity General Plan (RUKN) for 2024-2060 and the 10-year Electricity Supply Business Plan (RUPTL) for 2025-2034 (MEMR, 2024a, 2025). The long-term plan provides high-level projections with coarse national spatial details. targeting a sixfold increase in installed capacity to 442 GW while pursuing energy security and decarbonization through measures such as hydrogen, ammonia co-firing, and carbon capture. In contrast, the 10-year plan offers a detailed project-level plan across provinces, including both existing and planned generation and transmission assets, with 59 GW of new capacity to support annual demand growth of 5.3%. Beyond those planning documents, climate-related considerations have begun to appear in sectoral governance through disclosures under the Task Force on Climate-related Financial Disclosures (TCFD) framework (PLN, 2025).

*2.3 Climate Change Context*

Climate change has led to continued increases in ambient temperatures and sea level rise in Indonesia. Observed temperature records from Indonesia's meteorological stations show a clear warming trend since the 1981 baseline, with average temperatures increasing by about 0.8 °C and reaching their highest level in 2024 compared to the four-decade mean (BMKG, 2025). Average sea level trends increase 4,3 ± 0,4mm per year (BMKG, 2024). In addition, monsoon interaction and the El Niño–Southern Oscillation (ENSO) can further amplify temperature anomalies during dry periods, while La Niña is associated with increased precipitation, elevating the risks of flooding and landslides. These impacts are particularly significant in Indonesia, an equatorial archipelagic country with extensive mountainous terrain formed along the Pacific "Ring of Fire."

Recently, extreme weather events have caused severe impacts, such as the tropical cyclone that struck Aceh, Sumatera—an event rare for Indonesia—bringing intense rainfall and triggering floods and landslides (Tempo, 2025). Together, climate change and extreme weather phenomena create significant exposure across Indonesia's large and climatically diverse territory.

## 2.4 Climate-Related Exposure Across Indonesia's Power System Components

Climate change and extreme weather conditions expose Indonesia's power system, affecting generation, transmission, and electricity demand. The nature and manifestation of this exposure vary across locations and system components. A synthesis of established climate–power system interactions from the literature is provided in the Supplementary Material A. To support this overview, the evidence from Indonesia is discussed subsequently.

### 2.4.1 Generation assets

The major generation technologies in Indonesia's current and planned system are conventional steam-based power plants, including coal-fired units (MEMR, 2025), geothermal, some biomass, and future nuclear plants. Given Indonesia's archipelagic geography, most coal power plants utilize seawater cooling, which is sensitive to sea surface temperature and vulnerable to sea level rise, flooding, and jellyfish blooms (Handayani et al., 2019; PLN, 2025). Gas turbines, including open-cycle (OCGT) and combined-cycle (CCGT), represent the second major technology, providing load-following and peaking capacity for system reserves (MR 20/2020, 2020). However, their output declines with increasing ambient temperature, and they are also exposed to flooding risks due to land subsidence and sea level rise (Handayani et al., 2020; PLN, 2025; Center of Urban and Regional Resilience Research et al., 2025).

On smaller islands, diesel generators remain essential and are planned to be retrofitted. With some replaced by solar photovoltaic (PV) systems. Given Indonesia's large capacity expansion targets and declining costs, PV is expected to account for a substantial share of the system (MEMR, 2024a). However, PV efficiency is also known to be sensitive to rising ambient temperature (Dewi et al., 2019; Tarigan, 2019), and certain PV systems are exposed to flooding and fire risks (PLN, 2025). Moreover, Indonesia has significant potential for hydropower, wind, and ocean energy, but these are climate sensitive: hydropower declines during drought, and wind power may be vulnerable to extreme wind events. In general, Indonesia's power generation infrastructure is increasingly exposed to climate-induced derating and disruption.

### 2.4.2 Transmission assets

Indonesia operates approximately 71,886 km of high-voltage transmission lines (MEMR, 2025). High-voltage transmission systems operate at 66, 150, 275, and 500 kV, connecting generation assets and load centers across complex geography. Massive transmission systems increase exposure to climate change and weather-related disruptions (PLN, 2025). Flooding in Jakarta has repeatedly caused power outages in several substations (The Jakarta Post, 2020). Cyclone Seroja in East Nusa Tenggara brought heavy rain and landslides, disrupting electricity supply (ACE, 2021). A landslide in Pacitan damaged a four-circuit transmission tower serving a major coal-fired power plant (detik.com, 2024). Recently, extreme weather with a rare tropical cyclone in Aceh damaged five transmission towers due to flash floods and landslides (Tempo, 2025). Severe thunderstorms can damage transmission towers and substation equipment, and strong winds may push uncleared vegetation toward transmission lines, causing flashover or induced faults. Increased rainfall accelerates vegetation growth and maintenance needs, while prolonged dry seasons raise wildfire risk that exposes transmission assets to extreme heat.

### 2.4.3 Power Demand

Beyond generation and transmission, climate change also affects the power system through demand. In a growing economy, Indonesia also sees steady increases in electricity demand, partly driven by increased cooling needs (Atma et al., 2025; McNeil et al., 2019; MEMR, 2024a; Sherman et al., 2022). Air conditioning already accounts for a significant share of consumption, particularly in the residential and commercial sectors, making demand highly sensitive to rising temperatures. Higher temperatures increase cooling loads, a factor to be considered in power system planning.

## 3. Methods

Building on the previous section, this study develops an analytical framework to assess Indonesia's power system's vulnerability to climate change. We distinguish between two types of impact: "stressors", representing the primary, continuous pressures on system performance, and "shocks", representing additional discrete event-based hazards that cause disruption. This study focuses only on ambient temperature as the primary stressor and on sea level rise, flooding, and landslides as additional representative climate shocks, reflecting major hazards in Indonesia.

To assess both types of impacts, the method employs a spatial framework based on climate and geospatial data, together with empirical stressor functions and shock indices. It integrates bottom-up component-level assessment with a top-down spatial analysis to capture broader system-level interactions. A spatial approach is essential because both stress- and shock-related hazards are inherently locational, and shaped by geographic exposure and asset distribution. Technologies are generalized by type to provide a first-order approximate assessment rather than precise plant-level estimates. Nevertheless, the framework remains flexible to more detailed component-level asset parameterization as availability improves.

This study framework is illustrated in Fig. 2. Climatic, geographical, and hazard datasets, including elevation, temperature, sea-level rise, landslide, and flood maps, are used to assess the exposure of power system assets.

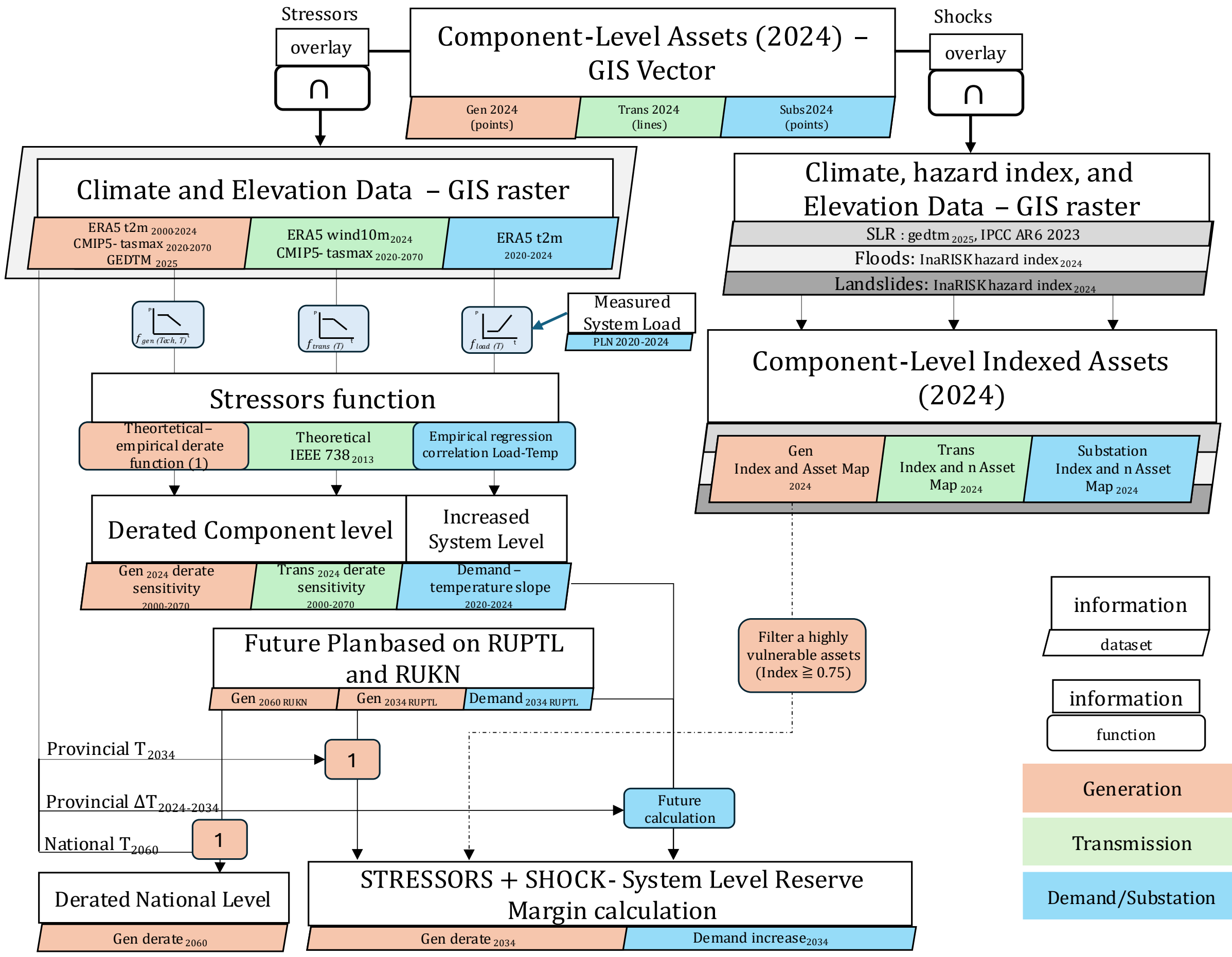

*Fig. 2. A framework used in this paper to assess stressors and shocks in Indonesia's power system.*

Historical climate data include 2-m air temperature (t2m) and 10-m wind speed (10m wind) from the ERA5 reanalysis dataset (C3S, 2018). ERA5 is selected for its hourly temporal resolution and ~0.25° spatial resolution (~25 × 25 km), which provides sufficient detail for a country as large as Indonesia. Future temperature projections use daily maximum near-surface air temperature (tasmax) from the HadGEM2 model of Coupled Model Intercomparison Project-5 (CMIP5) Southeast Asia (SEA-22) regional climate model (RCM) (Collins et al., 2011). RCM is used because it provides improved local geographical representation compared to the global climate model (GCM) and offers a similar grid resolution in 0.22° (~22 x 22 km). Elevation data are obtained from the Global Ensemble Digital Terrain Model (GEDTM) (Ho & Hengl, 2025), with a 1-arc-second (~110m) resolution.

Sea-level-rise projections are obtained from the IPCC AR6 assessment model (Garner et al., 2021), providing ensemble-based mean sea level estimates at ~1° (~100 x 100km) resolution up to year 2150 under multiple confidence levels and Shared Socioeconomic Pathways (SSP) scenarios, also analyzed in conjunction with elevation data. Landslide and flood hazard maps (~350m resolution) are obtained from the National Disaster Countermeasures Authority (BNPB) (BNPB, 2025).

### *3.1 Assessment of Climate Stressors*

**GIS vector assets**

This study uses component-level spatial data on generation, transmission, and substation from PLN in 2024, available in GIS GeoPackage format. In contrast, the 2034 10-year plan is available only at the provincial level, while the 2060 long-term plan offers only a single national-level projection.

Due to the confidential nature of detailed 2024 component-level data, only aggregated final results are presented. The asset derating code is publicly available, along with one sample GeoPackage asset provided only to illustrate the data structure. Projected generation capacity from the publicly available 10-year and long-term plans is included in the repository.

**Overlaying assets on the climate raster stressors**

Overlaying is a basic method for GIS-based climate vulnerability analysis (Kasthala et al., 2024). Component-level PLN asset data for 2024—represented as points for generation units and substations, and lines for transmission—are overlaid with climate and elevation rasters. Temperature and elevation values are extracted at each asset location. Temperature data include historical (t2m ERA5 for 2000 to 2024) and future CMIP5 projections (tasmax up to year 2070 under RCP 4.5 and 8.5). These are applied to the same 2024 generator asset configuration to assess component-level sensitivity across scenarios while holding installed capacity constant, thereby isolating temperature-driven derating effects.

For transmission assets, temperature exposure is evaluated at line vertices, using the maximum tasmax value to represent the most critical segment. To account for cooling effects, historical 10-m wind conditions for transmission analysis from 2024 ERA5 data are used. consistent with typical transmission line height in Indonesia. Wind conditions are characterized by the 10th and 90th percentiles (0.58 m/s and 3.37 m/s), representing low and high wind scenarios, as shown in the Supplementary Material C.

For demand analysis, substation point locations are overlaid with historical 2-m air temperature (t2m) time series for the period 2020–2024. The extracted temperature values are then aggregated to the system level using a weighted average across substations, aligning with the spatial resolution of the available electricity demand data. These aggregated temperature series are subsequently used in the regression analysis described later.

**Stressors function**

In the existing 2024 condition, after the necessary climate variables are overlaid onto each asset, generation and transmission units are assigned specific derating characteristics. Each asset is flagged by its technology type, which determines the appropriate derating function to apply. For consistency, all assets within the same technology (e.g., PV, coal-fired units) are assumed to share common derating parameters across Indonesia. The stressor functions applied in this study are retrieved from the literature, technical standards, and equipment specifications, and are generalized to ensure consistency across asset types. The applied stressor functions are summarized in Table 1.

*Table 1. Summary of stressor functions used in the framework. The detailed equations and explanations are available in Supplementary Material B.*

| *Elements* | *Function** |
|---:|:---:|
| *Steam: Coal Fired Power Plant* | ↓ Capacity 0.34 % every ↑ 1°C [a)] |
| *Steam: Nuclear* | ↓ Capacity 0.44 % every ↑ 1°C [b)] |
| *Gas: Open Cycle (OCGT)* | ↓ Capacity 0.6 % every ↑ 1°C [c) d)] |
| *Gas: Combine Cycle (CCGT)* | ↓ Capacity 0.45 % every ↑ 1°C [c) d)] |
| *Diesel* | *Derated capacity due to elevation and air temperature [e)]* |
| *PV* | *Derated capacity due to irradiance, air temperature [d)]* |
| *Transmission Lines* | *Thermal energy balance equation*<br>*IEEE standard 783™ - 2012 f)* |
| *Load Demand* | *Correlation: Daily energy (MWh) – Daily t2m (°C) linear regression, controlled with date, day type, and holidays.*<br>*Future projection: Based on annual temperature comparison, demand group, and correlation* |

** ↓ = decrease, ↑ = increase*

*a) (Petrakopoulou et al., 2020)*
*b) (Attia, 2015)*
*c) (Handayani et al., 2020)*
*d) (M. D. Bartos & Chester, 2015)*
*e) (Wärtsilä Diesel Oy, n.d.)*
*f) (IEEE, 2013)*

Based on Table 1, thermal generation output typically decreases by ~0.3–0.6% per 1 °C increase in ambient temperature, consistent with prior studies (Attia, 2015; M. D. Bartos & Chester, 2015; Burillo et al., 2019; Handayani et al., 2019, 2020; Petrakopoulou et al., 2020). Gas turbines are primarily affected by ambient air temperature, while steam-based generations are additionally sensitive to cooling-water temperature. Ambient air temperature is used as a proxy for seawater temperature based on their observed similarity (see Supplementary Material B). For PV, a simplified irradiance-temperature-based derating function is used (M. D. Bartos & Chester, 2015). The diesel generator derating is based on manufacturer specifications, linking power output to ambient temperature and site elevation. In all cases, derated output is evaluated relative to the original nameplate capacity, which serves as the baseline condition for generation assets.

The transmission derating uses a theoretical steady-state thermal energy balance formulation following IEEE Standard 738™ (IEEE, 2013), as applied by a previous study (M. Bartos et al., 2016). Transmission capacity is defined by ampacity, the maximum current a conductor can carry. Ampacity is determined by the balance between resistive heating and heat dissipation through convective and radiative cooling, including solar heat gain. It decreases with heating and increases with wind cooling. Derating is evaluated relative to the maximum ampacity, which serves as the baseline condition for transmission assets. The formulation is implemented within a spatial mapping approach. A representative conductor type is assumed per voltage level. For lines spanning multiple raster cells, each line is assigned the highest climate stress along its route, as overall performance is determined by the most constrained segment

Electricity demand is analyzed using an ordinary least squares (OLS) regression model (James et al., 2013), to estimate the relationship between daily maximum ERA5 ambient temperature and daily province-level electricity demand from PLN. This enables the estimation of temperature sensitivity at the aggregated transmission system level. The model temporal and calendar controls include year, month, day type, day name, and holiday indicators to isolate temperature effects. This provides an estimate of electricity demand sensitivity to temperature. The visuals of the regression dataset are available in the Supplementary Material B.

These estimated sensitivities are then used to project future electricity demand by adjusting the 2024 baseline demand using the sensitivity coefficient and the difference between future and historical temperatures. As residential and commercial demand are most sensitive to cooling, their share of total demand from the 10-year plan is applied to each year of the projection.

### 3.2 Assessment of Climate Shocks

Compared to stressors, climate shocks are treated as discrete disruptive hazards that act as additional sources of system risk. This study considers sea level rise, flooding, and landslides, applying a similar spatial overlay approach but incorporating different parameters such as elevation maps, projected sea level rise, and vulnerability indices. For the shock assessment, generator and substation point features are generalized into 200 × 200 m and 300 x 300 m polygon areas according to their capacity, while transmission assets are represented by their line geometry. For all asset types, the maximum raster shock values within the mentioned area are used to represent worst-case asset-level risk. This reflects the notion that a single high-risk segment can significantly influence overall component vulnerability.

For sea level rise (SLR), polygonal assets geometries are evaluated using a pixel-based overlay with elevation data. Each asset footprint is intersected with the underlying elevation raster, and inundation is identified at the pixel level. An asset is classified as exposed if at least one pixel within its geometry is projected to be inundated under the SLR scenario. Uncertainty is captured across multiple scenarios (SSP2-4.5, 3-7.0, 5-8.5) combined with percentile estimates(P50 and P95). To avoid misclassification of existing marine areas as newly inundated, elevation values ≤ 0 m from the GEDTM dataset are excluded, ensuring that only land areas above mean sea level are considered in the exposure assessment. The analysis focuses on 2060 to align with the net-zero target period. This method considers only SLR effects, excluding land subsidence and storm surges. Then, the SLR index for each province is calculated as the ratio of the number of inundated assets to the total number of assets.

For floods and landslides, polygonal assets are clipped with the shoreline map (Badan Informasi Geospasial (BIG), n.d.), and overlaid with the spatial BNPB InaRISK hazard index (BNPB, 2025). The index is a spatially explicit, map-based hazard assessment index derived from historical records, terrain analysis, and the official BNPB hazard classification framework, which assigns values between 0 and 1, where 0 indicates the lowest likelihood of occurrence, and 1 indicates the highest hazard-prone area.

Finally, the overlay outputs vulnerability indices and the number of exposed assets for each sea level rise, flood, and landslide map. For provincial and national summaries, asset-level values are aggregated using simple averaging.

### *3.3 Future Plan Analysis and Compound Stressors - Shock*

In addition to the component-level assessment of existing assets, this study evaluates how the 10-year and long-term plans respond to climate stressors and shocks. This assessment is necessary since neither plan explicitly accounts for temperature-driven generation derating or demand increases.

The 10-year plan, RUPTL, defines two planning scenarios—the Renewable Energy Base (REBase) and the Accelerated Renewable Energy Development (ARED). Both scenarios are evaluated for generation derating. Future provincial generation capacities are evaluated using the planned capacities from the 10-year plan and the corresponding provincial temperature projections. Future electricity demand is projected using the earlier demand–temperature relationship.

To capture the combined effects of stressors and shocks, system resilience is assessed through system adequacy measured using the "reserve margin". The reserve margin is calculated for 2034 under original and temperature-stressed conditions. The REBase scenario is highlighted because it represents a planning configuration with lower installed capacity and reserve margins, enabling a clearer assessment of how climate-related effects influence system adequacy. Shock impacts are incorporated by identifying assets in areas with high hazard index values (InaRISK > 0.75), and removing their corresponding generation capacity from the system calculation. This additional capacity reduction further lowers the reserve margin and illustrates the compounding effects of climate shocks and stressors.

## 4. Results

This section presents the results for stressors, shocks, and their combined effects. The national-level aggregation of those is summarized in Table 2, which shows changes in generation, transmission, and demand under temperature stress, along with vulnerability levels across different shock hazards. Detailed explanations are provided in the following subsections.

*Table 2. Overview of stressor and shock impacts across power system components.*

| Component | Stressors | Shocks (0<index<1)* |
|---|---|---|
| Generation | ↓ 3-5 % capacity derate (^existing , ^plan–2034+2060) | SLR(^existing): 0.037<br>Floods(^existing-an): 0.357<br>Landslides(^existing-an): 0.268 |
| Transmission | ↓ capacity up to 6% in high wind (^existing)<br>↓ capacity up to 33% in low wind (^existing) | SLR(^existing): 0.109<br>Floods(^existing-an): 0.849<br>Landslides(^existing-an): 0.545 |
| Demand/Substations | Increasing Energy (^hist, ^plan-2034)<br>↑ 2–11% every ↑ 1 °C | SLR(^existing): 0.010<br>Floods(^existing-an): 0.391<br>Landslides(^existing-an): 0.164 |
| Reserve Margin | Decreasing Reserve Margin (^plan-2034), ↓ 6-36 percentage points (pp) relative across systems<br>Jamali: 47 % → 27% (20pp)<br>Sumatera: 79% → 53% (22pp)<br>Kalimantan System: 61% → 24% (36pp)<br>Northern Sulawesi: 62% → 53% (9pp)<br>Southern Sulawesi: 61% → 38% (23pp) | |

*InaRISK index (0–1): hazard intensity from BNPB spatial maps, assessed for the 2024 existing asset configuration.
*SLR index (0–1): share of 2024 existing assets exposed to inundation under the respective sea-level rise scenario in 2060.
*A bigger index shows a more vulnerable condition.
^existing = 2024 asset stressed up to 2060 climate.
^existing-an = 2024 asset with analytical-historical BNPB InaRISK hazard map.
^plan = Future planned capacity from 10-year plan and long-term plan.
^hist = Historical data from 2020-2024.

### *4.1 Impacts of Climate Stressors*

#### *4.1.1 Generation*

A straightforward way the impact of temperature on generation is to analyze the long-term plan, RUKN. This plan provides a coarse national-level representation of the projected 2060 energy mix. As shown in subplot (b) of Fig. 3, the worst-case RCP 8.5 temperature-related derating is estimated at 2.83% of total capacity, resulting in losses of 12.5 GW. Although these percentages appear small, the absolute losses are substantial and cannot be overlooked.

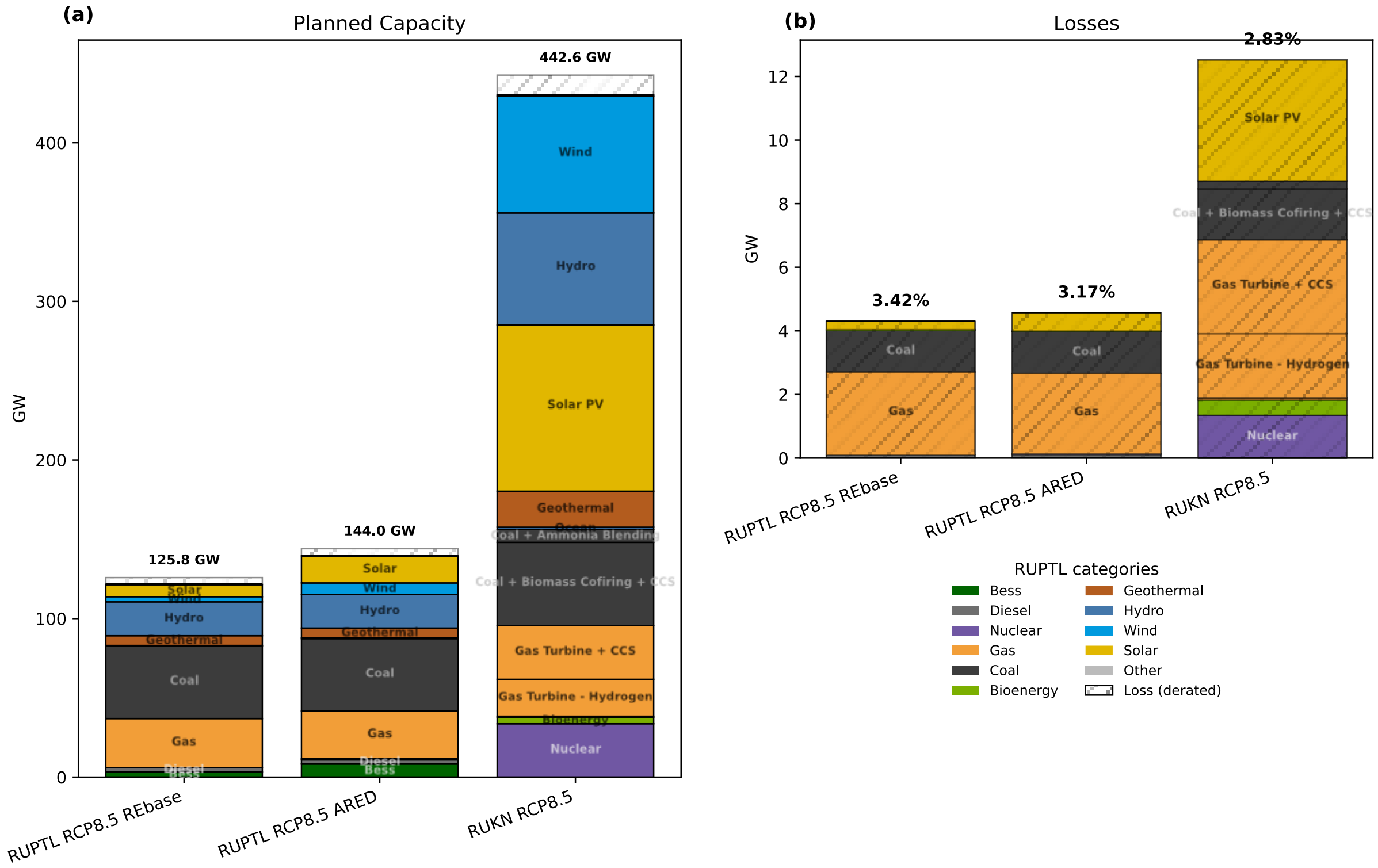

*Fig. 3. Comparison of temperature stress impacts under the long-term (RUKN 2060) and 10-year (RUPTL 2034) plans. Plot (a) shows the total planned installed capacity, and subplot (b) shows the temperature-based loss portion.*

Comparable to the national analysis in 2060, the provincial analysis is available in the 10-year plan. This helps confirm the provincial-level detail on assets and temperature. Subplot (b) of Fig.3 also shows the derate for 2034, around 3.17%-3.42%, equal to 4.2-4.3GW. Gas turbines and their variants (CCGT and OCGT) account for more than half of the total derating, with gas and coal together responsible for about 80% of the total capacity loss. Photovoltaic (PV) also experiences noticeable impacts due to its initial share of the system, whereas nuclear power and bioenergy are affected to a lesser degree.

To further examine the impact at the component level, the GIS asset data in the existing 2024 condition is evaluated. Fig. 4 presents the national aggregation of total derated existing capacity across years and scenarios. The RCP8.5 scenario, representing the worst-case pathway, returns the highest loss. In 2024, the average national derating loss is ~ 3.5%, corresponding to ~ 2,500 MW of unavailable capacity from Indonesia's total installed capacity of ~ 72,300 MW. This loss is projected to increase further, reaching around 5% by the 2060s' sensitivity analysis using the same generation asset. A noticeable bias is also observed between the ERA5 and CMIP5 datasets for the same year. In 2024, the average usable capacity derived from ERA5 is about 96.5%, whereas the CMIP5-based projection shows 95.3%. This is because CMIP5 temperatures were not bias-corrected against ERA5, so the differences between the two datasets reflect climate model bias rather than actual temperature conditions.

The derated composition of each type of power plant is illustrated in Fig. 4. The results show that gas-based power plants, whether CCGT, OCGT, or Gas Engine, consistently represent the largest share of total derating loss up to more than 60%, followed by coal plants, whose contribution increases from 21.8% in 2024 to around 27.4–28% in 2060. Diesel power plants also exhibit noticeable derating. Gas technologies experience a greater decline in performance than coal due to their higher sensitivity to ambient temperature. Meanwhile, Biomass, Biogas, and Solar PV (including floating) contribute only marginally to the total loss at the national scale, reflecting their relatively small installed capacity shares.

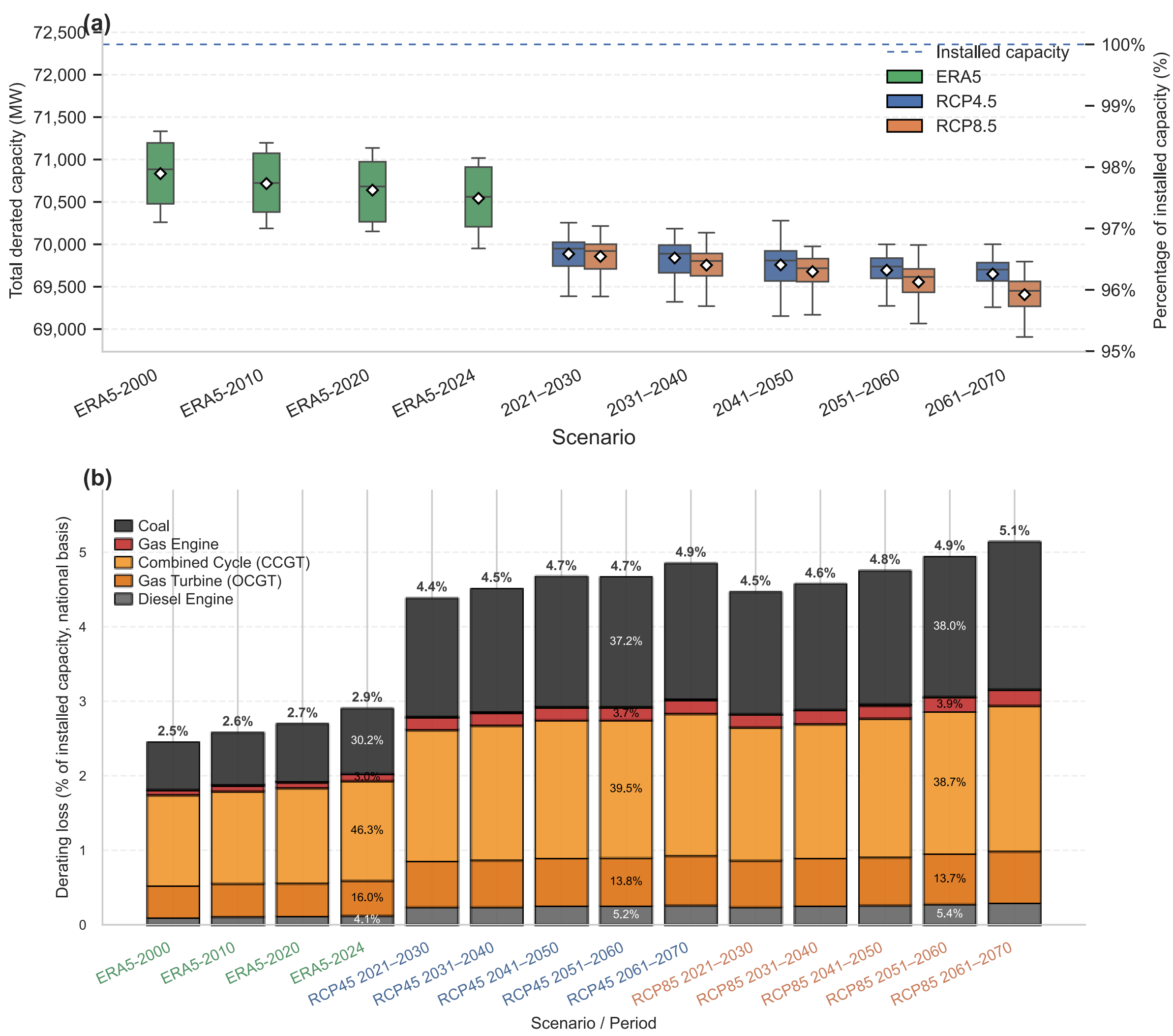

*Fig. 4. (a)Total derated generation capacity under historical ERA5 and future RCP scenarios (4.5 and 8.5), calculated using the same generation asset configuration from 2024. (b) The corresponding proportion of the loss.*

### *4.1.2 Transmission*

Temperature rise also stresses transmission lines. Fig. 5 presents the provincial transmission derating under different temperature and wind conditions. Overall, higher ambient temperatures lead to modest reductions in transmission capacity; however, wind speed emerges as the dominant moderating factor. Under high wind conditions (3.37 m/s), derating remains low across all provinces. In contrast, low wind conditions (0.58 m/s) substantially amplify thermal stress. Under these conditions, derating reaches up to 33% in Central Sulawesi and approximately 10% in the Riau Islands. Increased wind speeds significantly mitigate this effect, reducing the maximum derating in Central Sulawesi to around 8%. The figure indicates that wind variability and absolute provincial temperature, rather than long-term temperature trends, primarily drive operational transmission derating differences across provinces.

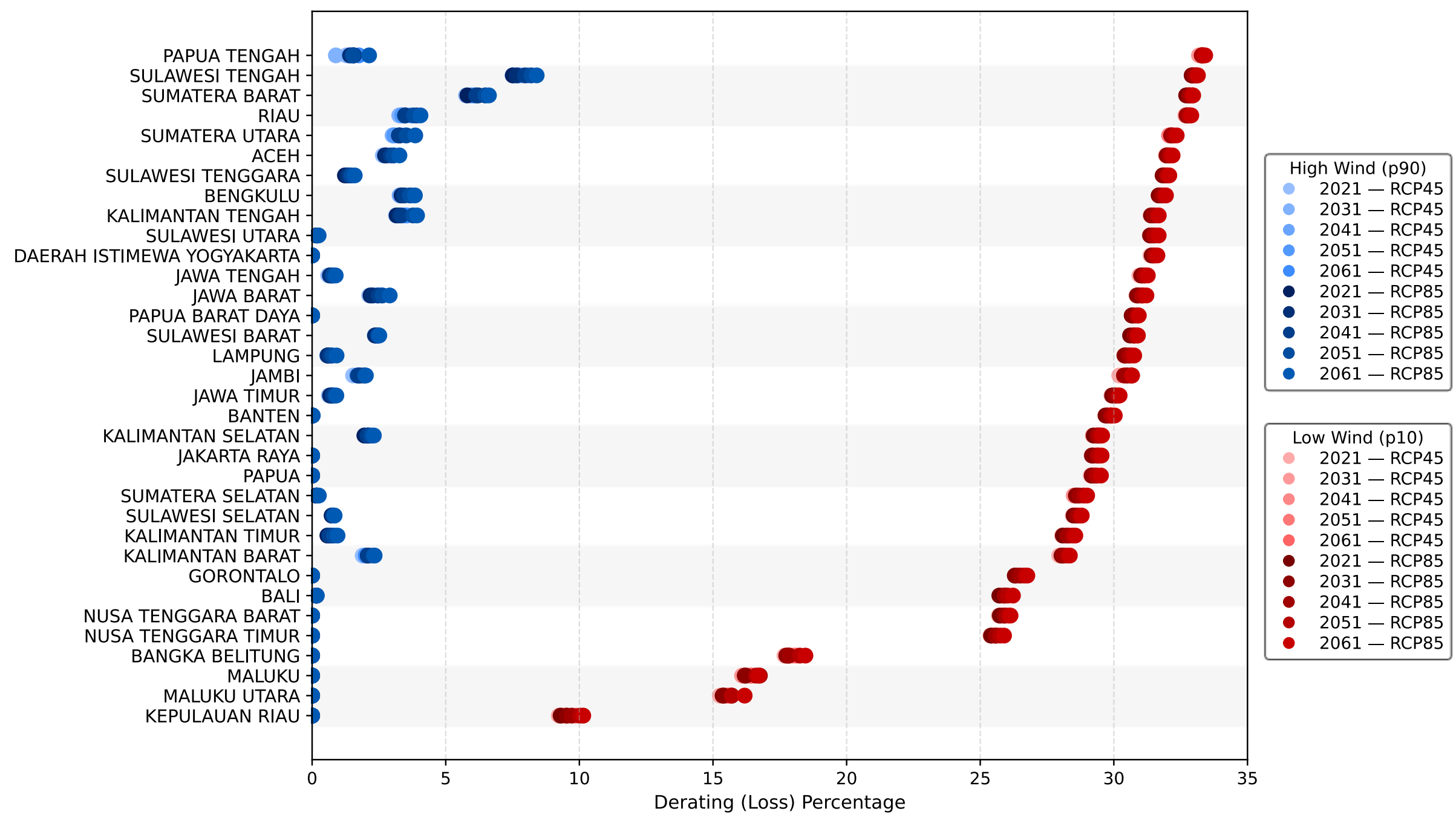

*Fig. 5. Per-province transmission line derating percentage under low and high wind speeds across the decades 2021-2061. A higher derating percentage means a higher reduced capacity.*

### *4.1.3 Demand*

The final element stressed by temperature is electricity demand. Using the regression analysis, Fig. 6 summarizes the correlation results. Most areas exhibit a strong positive correlation, with Bali reaching the highest value of 0.95. In contrast, Sulawesi Utara, Sumatera Barat, and Bangka show moderate correlations of around 0.5–0.6. Visual on the provincial-level time-series plots presented in Supplementary Material B. From that visual, these three provinces exhibit relatively stagnant demand growth in recent years, which likely contributes to the weaker temperature–demand correlation. In contrast, provinces with more pronounced demand growth tend to show stronger correlations

Fig. 6 indicates that nearly all areas have a positive percentage change per degree Celsius, except one case, Bangka, which has an anomaly in industrial growth (BPS, 2021). Bali shows the highest increase at approximately 10.9%, while Jawa Barat records the lowest at 2%. This implies that a 1°C rise in temperature is associated with an estimated 10.9% increase in demand in Bali, compared to only 2% in Jawa Barat. For absolute demand, DKI Jakarta, Bali, and Jawa Timur experience the largest annual demand increases, on the order of 3.9–2.5 TWh.

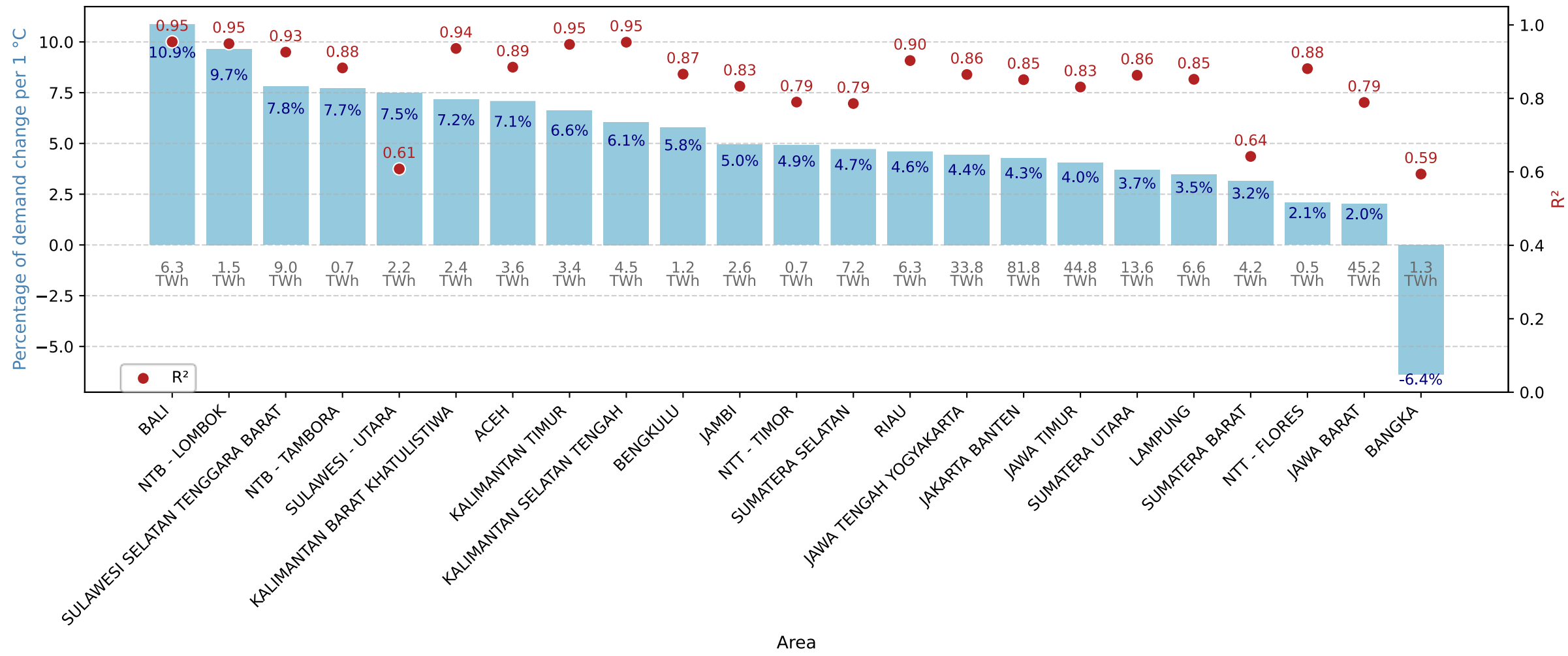

*Fig. 6 Correlation summary between ERA5 daily maximum temperature and daily area-level demand. The percentage shows the slope of demand change for every 1-degree Celsius change. The energy value in TWh in each bar shows the yearly average demand in 2021-2024.*

Bangka, which shows a -6% slope, may have resulted from an anomaly in demand growth and industrial activity, which will be discussed further as a model limitation in Section 5. Given the regression model's inability to capture unusual demand and industrial growth, Bangka will be excluded from the subsequent analysis. The least provincial growth of future demand will instead adopt the lowest credible slope observed in Jawa Barat. Using the lowest credible slope preserves conservative estimation while avoiding unrealistic zero or negative slopes.

Taken together, the temperature stressor of Generation-Transmission-Demand across all Indonesian provinces is delivered in Fig. 7. For extremes, Bali shows the highest demand-temperature slope, and DKI Jakarta shows the highest derated portion of generation. Transmission derate shows a similar derate over provinces.

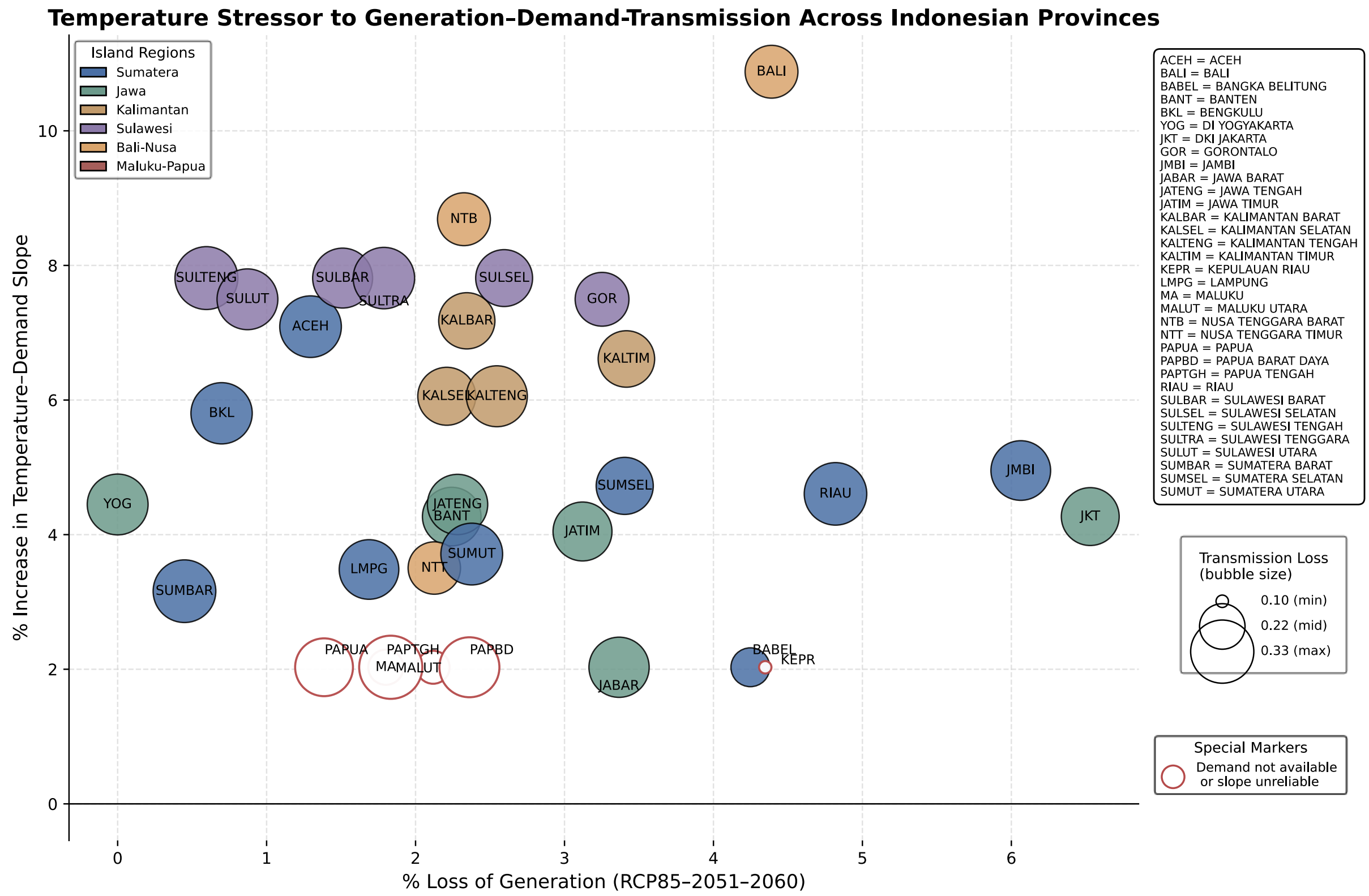

*Fig. 7. Temperature stressors on generation, transmission, and demand across Indonesia's provinces.*

## *4.2 Impact of Climate Shocks*

### *4.2.1. Sea Level Rise (SLR)*

The national impact of SLR and the detailed count of assets at risk in the year 2030 up to 2060 are shown in Fig. 8. The figure shows the inundated assets calculated by comparing asset elevations with the projected sea level. In 2030, the method indicates no inundation under the lowest scenario, increasing up to 26 generation assets, 20 transmission lines, and 2 substations under the highest scenario. By 2060, between 39 and 75 generations, 32 and 50 transmission lines, and 2 – to 4 substations are projected to be inundated across the lowest to highest scenarios. This reflects the inherent uncertainties in the assessment, while also capturing known vulnerable assets, such as the Semarang–Demak transmission corridor, which has been reported to be at risk of coastal inundation. Further discussion about this finding will be delivered in the next section. In summary, the increasing number of inundated assets in the far future indicates that assets are slowly becoming at risk of inundation.

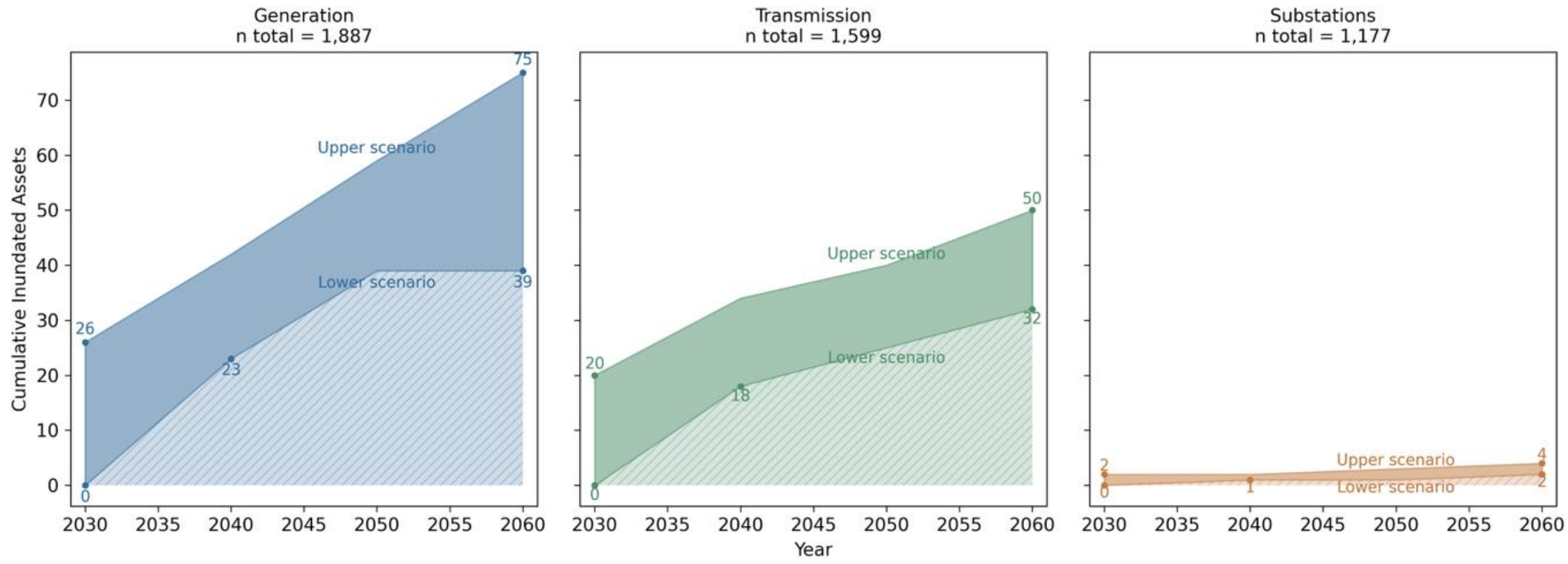

*Fig. 8. National cumulative exposure of electricity infrastructure to sea level rise (SLR). The upper envelope represents the highest projected count of inundated assets across SSP2-4.5, SSP3-7.0, and SSP5-8.5 under the high-end estimate (95th percentile), while the lower envelope represents the corresponding median projections (50th percentile) across the same scenarios.*

### *4.2.2 Floods and Landslides*

While SLR represents long-term future risk, flood and landslide hazards represent immediate risks. We use the BNPB vulnerability index, calculated based on current condition, and overlay it with the 2024 component-level power asset dataset to identify the corresponding vulnerability index for each asset.

Fig. 9 shows the provincial aggregate of the flood and landslide vulnerability. Transmission assets appear most vulnerable across all shock hazards. This is likely because transmission lines are extensive, critical, and connect generation plants to substations, resulting in a large number of assets exposed. Conceptually, flood and landslide characteristics differ significantly. While transmission towers may face flood risk, short-duration flooding is generally less disruptive for these structures. Landslides, however, pose a serious threat, as evidence shows they frequently occur in Indonesia and can severely disrupt electricity delivery. Overall, Fig. 9 indicates that power system assets remain vulnerable to climate shocks.

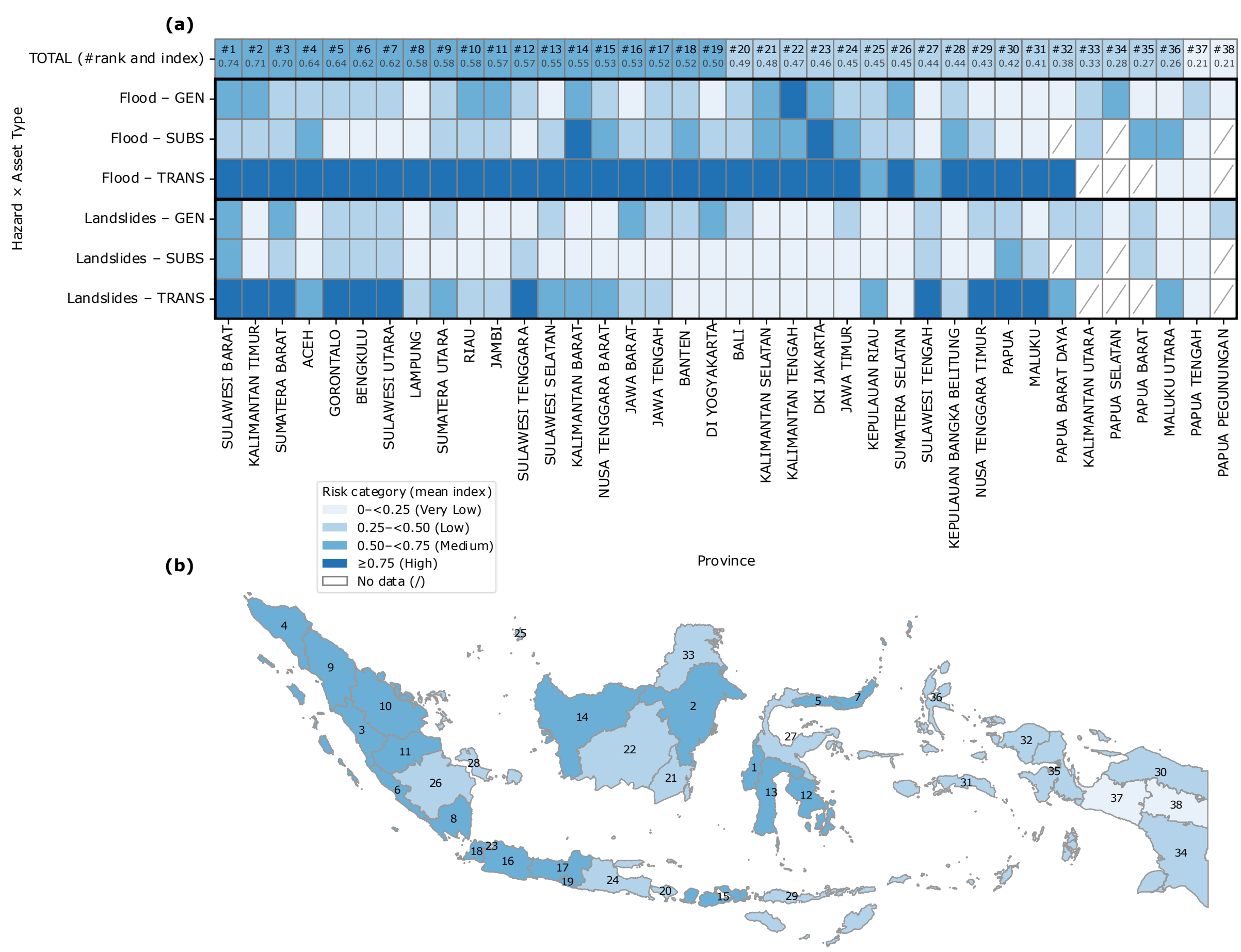

*Fig. 9. Provincial flood and landslide vulnerability index for power system assets (higher values indicate greater vulnerability). (a) Vulnerability matrix by province and asset type; the top row shows the asset-count-weighted total index and rank (diagonal marks indicate missing data). (b) Choropleth map of the same index with provincial ranks.*

## *4.3 Combination of Climate Stressors and Shocks Affecting the System's Reserve Margin*

In the combined assessment of stressors and shocks, provincial stressed generation capacity and demand are further adjusted by excluding assets in high-shock-index areas. We analyzed the reserve margin for 2034 from the 10-year plan for the assessment. Fig. 10 presents the reserve margin under the combined impacts of temperature-related stressors and flood/landslide shocks on supply and demand in 2034 in major systems in Indonesia. The reserve margins decline substantially. In Jawa-Madura-Bali (Jamali) system, the reserve margin decreases from 47.3% to 26.5%. Reductions also appear in Sumatra (79% → 53.1%), Kalimantan Interconnection (60.9% → 24.2%), the Northern part of Sulawesi (61.6% → 52.9%), and the Southern part of Sulawesi (60.5% → 38.7%). That means the erosion varied from 9 to 36 percentage points (pp). Kalimantan experiences the largest decline (36 pp), while Northern Sulawesi shows the smallest reduction (8.7 pp). The largest system, Java–Madura–Bali, records a substantial erosion of 20.8 pp.

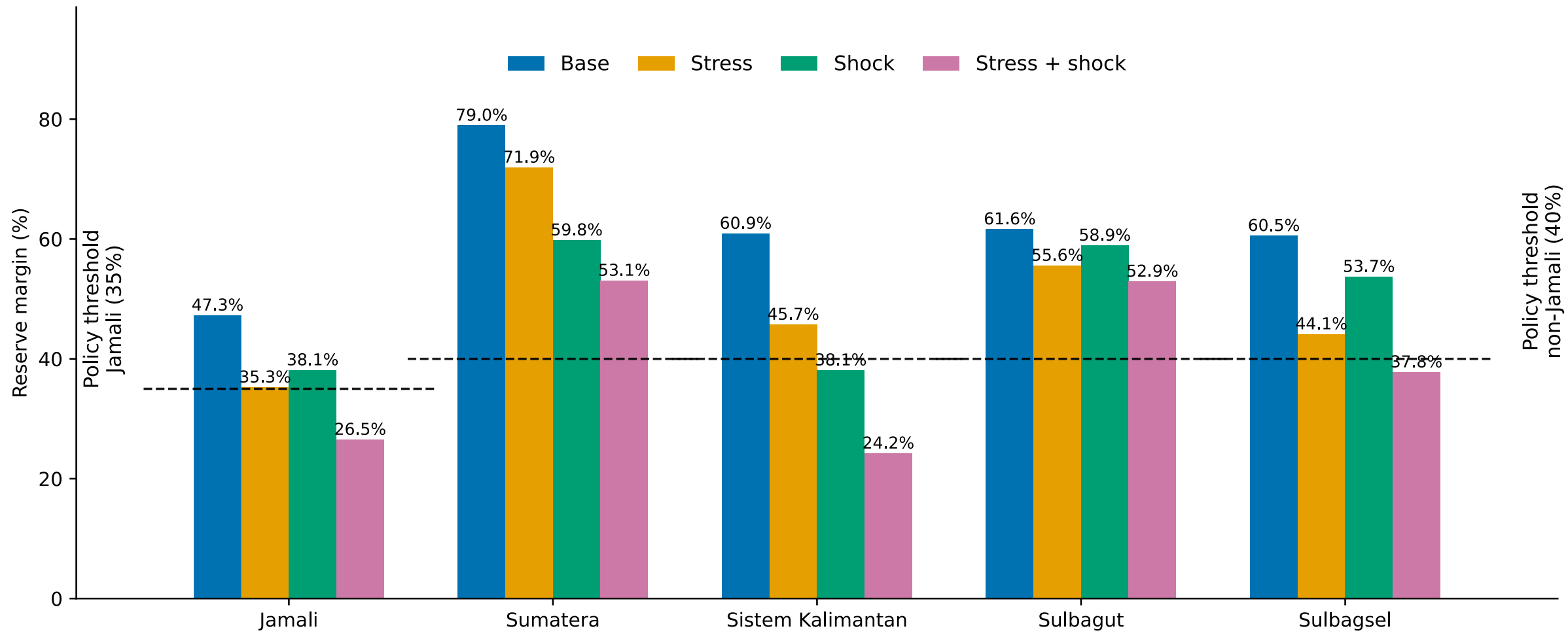

*Fig. 10 Reserve margin decline under temperature stress and compound climate shocks across major Indonesian power systems. The base data was obtained from the 10-year plan for year 2034. Bars show reserve margins under baseline conditions, temperature stress, shock exposure (floods and landslides with risk ≥ 0.75), and their combined effects.*

These reserve margins can be compared with the 10-year plan policy threshold used to plan, 35% for Jamali and 40% for outer-Jamali. Under only stressors or shocks, the system's reserve margin is still above the threshold; however, under combined conditions, the system's reserve margin goes beneath the threshold. While Fig. 10 shows a high-level illustration of the system-level reserve margin, more details on supply and demand in each province are provided in the Supplementary Material D.

## 5. Discussion

The results indicate that climate change poses a systemic challenge to Indonesia's power system. Temperature stress simultaneously degrades generation and transmission capacity while increasing electricity demand, tightening reserve margins. Not only do climate stressors affect individual components in isolation, but they also interact across generation, transmission, and demand, amplifying their combined impact on system adequacy. When compounded by acute climate shocks—such as sea-level rise, flooding, and landslides—these interactions further constrain reserve margins and reduce the system's ability to absorb future uncertainties, revealing a structural vulnerability in long-term power system planning.

### *5.1 Interpreting the Results: Implications for System Adequacy and Resilience*

These findings indicate that climate change undermines system adequacy, which refers to the ability of the power system to reliably meet electricity demand under normal and peak conditions. Climate stressors not only reduce effective generation and transmission capacity but also simultaneously increase electricity demand. As a result, a growing share of future investments may be required merely to offset climate-related losses, rather than to expand usable system capacity.

The plan, both the 10-year plan, RUPTL, and the long-term plan, RUKN, continues to rely on temperature-sensitive thermal generation—particularly gas and steam turbines, including those mitigated through options such as CCS/CCUS or hydrogen—to ensure system stability. At the same time, new technologies such as nuclear and solar PV are not immune to climate impacts. Nuclear remains subject to thermal limitations due to its reliance on steam turbines, while solar PV contributes limited firm capacity because of variability and the absence of nighttime output. This continued reliance implies that future adequacy risks are increasingly shaped by climate exposure rather than by installed capacity alone.

One important finding in the 2060 plan is that the 12 GW loss in national generation capacity due to higher temperatures is roughly equal to one-fifth of Indonesia's current capacity, meaning a significant portion of future investments may go just to offset climate-related losses rather than expand supply. Several regions experienced extreme loss in capacity, and other regions experienced extreme demand increase. This spatial divergence—where some regions face pronounced capacity losses while others experience rapid demand growth—further complicates system-wide adequacy planning. This finding challenges the implicit assumption in long-term planning that installed capacity can be treated as fully available under future climate conditions.

Climate shocks further compound these challenges by introducing localized, component-level disruptions that can escalate into system-wide stressors. The shocks in the generation complex directly disconnect the generation. Meanwhile, shocks in transmission infrastructure, such as a segment or tower due to landslides, can disconnect entire lines and, in some cases, multiple circuits simultaneously, and disconnect some power plant complexes. The 10-year plan stated that a reserve margin of 35% is needed in Indonesia, taking into account loss-of-load probability, auxiliaries, plant derating due to aging, energy availability, etc. When shock is combined with temperature-related stress, reserve margins in Indonesia's most complex system, Jawa, decline from around 40% to 26.5%, falling below the 35% planning threshold specified in the 10-year plan. This erosion of adequacy buffers indicates that climate stressors and shocks jointly push the system beyond conditions assumed in long-term planning.

The results of these stressors and shocks studies show that climate adaptation is becoming a prerequisite for effective power system expansion. Rather than translating directly into improved adequacy, a substantial share of planned capacity expansion may be absorbed by climate-induced derating and demand amplification, requiring continued investment simply to maintain existing reserve margins. This finding is consistent with previous studies that highlight increasing investment needs to offset climate impacts on power systems (Ralston Fonseca et al., 2021).

### 5.2 Contribution to Existing Climate–Energy Studies

The findings of this study help situate Indonesia's power system vulnerability within both national and global climate–energy literature. Previous work in Indonesia has established the relevance of climate change for the power sector, primarily through qualitative assessments, an energy optimization plan in the Jawa-Bali system, and policy-oriented reporting, including early vulnerability mapping and disclosures (Handayani et al., 2019, 2020; PLN, 2025). While these studies provide important contextual insights, the quantitative propagation of climate impacts across system components and their spatial variation at the national scale remains insufficiently explored.

Previous studies on climate impacts in Indonesia's power sector have primarily examined individual components, such as generation technologies, transmission outages, or electricity demand (Dewi et al., 2019; Jessen et al., 2022a; Tarigan, 2019; Veanti et al., 2022). Other work has addressed climate stressors and shocks separately, without explicitly considering their combined effects across generation, transmission, and demand (Handayani et al., 2020; Jessen et al., 2022a). The results presented here indicate that climate impacts become most evident at the system level, where generation derating, transmission constraints, and demand amplification interact to undermine system adequacy. By examining all major power system components at both component and system levels, this study reveals how climate stressors and shocks propagate across the system. In particular, component-level shocks may translate into system-level stress, while gradual component-level stressors may escalate into broader system-wide impacts.

Globally, vulnerability studies of power systems have long recognized temperature stress and extreme events as key drivers of power system vulnerability, particularly for thermal generation and transmission networks (Attia, 2015; M. D. Bartos & Chester, 2015; Burillo et al., 2019; Ke et al., 2016; Panteli & Mancarella, 2015; Petrakopoulou et al., 2020). Building on this body of work, the results presented here illustrate how these mechanisms operate within Indonesia's specific climatic and infrastructural context. By examining the effects of gradual temperature stress and acute climate shocks within a spatial framework, the analysis shows how theoretical thermal limits and localized disruptions can combine to create binding system-level constraints under future planning assumptions, particularly in geographically complex, climate-exposed power systems.

*5.3 Policy Implications for Indonesia's Power System Development*

Indonesia's power sector planning has focused on decarbonization objectives and has recently begun to acknowledge the need for climate change adaptation. Current regulatory frameworks, including the 10-year plan, the long-term plan, and the grid code, primarily address system resilience through conventional adequacy metrics, such as reserve margin and contingency criteria. However, these frameworks have not explicitly incorporated climate-driven derating, demand amplification, or compound stressor–shock effects, indicating that climate adaptation remains a secondary consideration in long-term system planning. This challenge is further compounded by existing policy dynamics, which continue to prioritize meeting growing electricity demand through conventional supply expansion, often at the expense of accelerating renewable energy deployment (Massagony et al., 2025).

Reliability in Indonesia's power system is traditionally assessed using reserve margins and contingency criteria, typically formulated as an N-1 criterion, which requires the system to withstand the failure of a single major component, such as a single line disturbance. However, climate-related hazards such as extreme heat, flooding, or landslides may affect multiple assets simultaneously. In transmission systems, for example, a landslide can damage an entire tower comprising multiple lines. Under such conditions, system disturbances may resemble N-k rather than N-1, where k is the number of lines. The results of this study, therefore, suggest that reliability assessments should increasingly consider climate-informed worst-case conditions (Panteli & Mancarella, 2015).

These findings have direct implications for technology choices and investment strategies in Indonesia's power system. Climate adaptation may require a combination of measures, including the deployment of more climate-resilient technologies, targeted upgrades to existing assets—such as improved cooling systems for thermal power plants—increased demand-side flexibility through smart grids and demand response, reduced hazard exposure in future asset siting, as well as network reinforcement and asset hardening. From an operational perspective, adaptation may also involve more climate-aware utilization of existing infrastructure, for example, through approaches such as Dynamic Line Rating, which can help reflect weather-dependent transmission capacity rather than relying on static assumptions. Rather than relying on a single solution, effective adaptation is likely to involve coordinated investments across generation, transmission, and demand-side infrastructure.

While these adaptation measures imply higher operational costs and additional investment requirements, their purpose is to safeguard system reliability under future climate conditions and to reduce the risk of large-scale disruptions, such as widespread blackouts. From a policy perspective, this highlights the need to align planning, regulatory, and investment frameworks with climate-resilient power system development. As climate adaptation begins to enter power sector planning, these results suggest it may be critical for safeguarding the effectiveness of future investments under climate change.

### *5.4 Limitations and Future Works*

The vulnerability assessment in this study is subject to several limitations that should be considered. First, uncertainties in climate data resolution and accuracy affect the output. Low-resolution climate models may underestimate temperature extremes (Chang et al., 2025), while sea-level-rise exposure is assessed using a static elevation-based approach that excludes land subsidence, storm surge, tidal dynamics, and local adaptation measures, potentially leading to conservative or incomplete estimates of coastal risk. Elevation-based sea-level-rise assessments are also known to overestimate exposure due to the exclusion of coastal dynamics and protection measures (Seeger & Minderhoud, 2026) reflecting a common challenge in large-scale screening approaches. The accuracy of climate and geographical models, such as the GEDTM elevation dataset used and future temperature, wind, and sea-level-rise projections, plays a vital role. Therefore, improving future assessments requires more accurate measurement-based data through collaboration with local authorities and infrastructure owners.

Second, asset representation and modelling assumptions are necessarily simplified. Temperature-related derating relies on generalized parameters and captures only direct thermal effects, without explicitly accounting for aging, technology-specific cooling configurations, or other operational constraints. Similarly, the PV model does not account for wind effects or long-term degradation, and transmission impacts are assessed using static ampacity assumptions rather than dynamic power-stability constraints. Demand–temperature relationships are estimated using regression models that capture temporal and calendar effects but do not explicitly represent structural changes in economic activity or other non-climatic drivers.

Third, the analysis does not include an explicit network or optimization model. Transmission stress and reserve margin impacts are assessed using asset-level data and data-driven calculations rather than power-flow or system-optimization simulations. As a result, spatial bottlenecks, network constraints, and correlated outages are not explicitly represented, and more detailed network-based simulations would be required to capture these effects.

Fourth, this study limits its scope to documenting and combining temperature-related stressors with flood- and landslide-related shocks, without examining possible systemic relationships between these hazards. While these phenomena do not necessarily coincide at the same location or time—since floods and landslides often occur during periods of lower temperatures—the large spatial scale of the national power system means that such conditions can plausibly occur simultaneously across different regions, warranting their joint consideration as system-level extremes.

Future work could take several directions. First, the results show concurrent climate constraints across generation, transmission, and demand, which cannot be fully interpreted using an asset-level list alone. Integrating system-level modelling with the spatial vulnerability results presented here could help represent locational reserve margins, spatial bottlenecks, and operational constraints more explicitly. Second, the results show that key inputs—particularly climate data and demand projections—strongly shape vulnerability outcomes under future conditions characterized by high uncertainty. Future research could therefore emphasize uncertainty-aware approaches to assess the robustness of system-level insights. Finally, further work is needed to examine how climate-informed vulnerability and adequacy insights could be incorporated into existing planning processes, including long-term expansion planning, reliability standards, and regulatory frameworks such as the 10-year plan, long-term plans, and the Grid Code.

## 6. Conclusions

This study examined Indonesia's vulnerability to climate-related stressors and shocks to understand how their combined effects may undermine power system resilience. Using a spatial, asset-level assessment framework, the analysis captures climate-driven vulnerabilities across generation, transmission, substations, and electricity demand at the national scale. While the framework does not rely on full network simulations, it provides a first-order, system-wide approximation of where and how climate impacts may align across different power system components. It also offers flexibility to add component-level details to every asset.

The results indicate that temperature-related stress acts as a dominant and pervasive driver of vulnerability, simultaneously reducing effective generation and transmission capacity while amplifying electricity demand. When combined with climate shocks such as sea-level rise, flooding, and landslides, these stressors further erode reserve margins and challenge conventional assumptions about system adequacy in several major Indonesian power systems.

Taken together, these findings suggest that climate change is not a peripheral risk to Indonesia's power sector, but a systemic constraint that increasingly shapes the effectiveness of capacity expansion and planning. Incorporating asset-level climate vulnerability into planning and operational decision-making can therefore provide critical insight for safeguarding future investments and supporting a resilient energy transition under changing climate conditions.

**Author contributions**

H.A.: Conceptualization, Methodology, Software, Validation, Formal Analysis, Writing – Original Draft, Data Curation, Visualization, Funding Acquisition.
N.G.: Conceptualization, Formal analysis, Writing – Review & Editing, Supervision.
S.P.: Conceptualization, Formal analysis, Writing – Review & Editing, Supervision.
I.N.: Conceptualization, Formal analysis, Writing – Review & Editing, Supervision.

**Conflict of interest**

The authors declare that they have no known competing financial interests or personal relationships that could have appeared to influence the work reported in this paper.

**Funding**

This research was funded by the Indonesian Endowment Fund for Education (LPDP), Ministry of Finance of the Republic of Indonesia, under grant number 0005412/ENE/D/19/lpdp2023

**Data availability**

The code supporting this study is available at https://doi.org/10.4121/0dcee8e7-0723-4365-9d55-4fc1d4ba2feb.

**Supplementary Material**

Supplementary material includes detailed formulations of the derating functions for generation, transmission, and demand, additional spatial analysis results, and extended figures supporting the stressor and shock assessments.

**Declaration of generative AI and AI-assisted technologies in the manuscript preparation process.**

During the preparation of this work, the authors used generative AI tools to assist in developing code for data processing and visualization. The authors reviewed, validated, and adapted all outputs as needed and take full responsibility for the accuracy and integrity of the results and the content of the published article.

1

# Supplementary Material

## Table of Contents

## A. Climate Change Impact To Power System Elements

Table A. 1 consolidates evidence from the literature on the sensitivity of power system assets to key climate variables. The impacts are organized by component type and hazard category, including temperature stress, flooding and sea level rise (SLR), precipitation effects, and landslides. For each technology, the table distinguishes between operational impacts (e.g., capacity derating, fuel quality degradation, altered runoff) and structural risks. (Azevedo De Almeida & Mostafavi, 2016; Ebinger & Vergara, 2011; Gonçalves et al., 2024; Kostevica & Dzikevics, 2023; Ma et al., 2022; Panteli & Mancarella, 2015; Russo et al., 2023; Wang et al., 2022)

*Table A. 1. Impact of temperature, flooding, and landslides on power system components. The notation ⊠ means a type of disruptive impact.*

| *Elements* | Temperature | Flooding / Sea Level Rise (SLR) / Precipitation | Landslides |
|---|---|---|---|
| | Generation | | |
| *Coal* | High Temp: ↓ capacity | High precipitation: Coal quality degradation.<br>Floods: increase demand in water drainage pump<br>⊠ SLR: Coastal asset risk | ⊠ Structural Damage |
| *Gas Turbine* | High Temp: ↓ capacity | ⊠ SLR: Coastal asset risk | ⊠ Structural Damage |
| *Biomass* | Crop growth affected | Crop growth affected | ⊠ Structural Damage |
| *Nuclear* | High Temp: ↓ capacity | Floods: increase demand in water drainage pump<br>SLR: Coastal asset risk | ⊠ Structural Damage |
| *Hydro* | Runoff River Changes | Floods: Dam spillage | ⊠ Structural Damage |
| *Solar Panel / PV* | High Temp: ↓ capacity, | High precipitation: Coal quality degradation. | ⊠ Structural Damage |

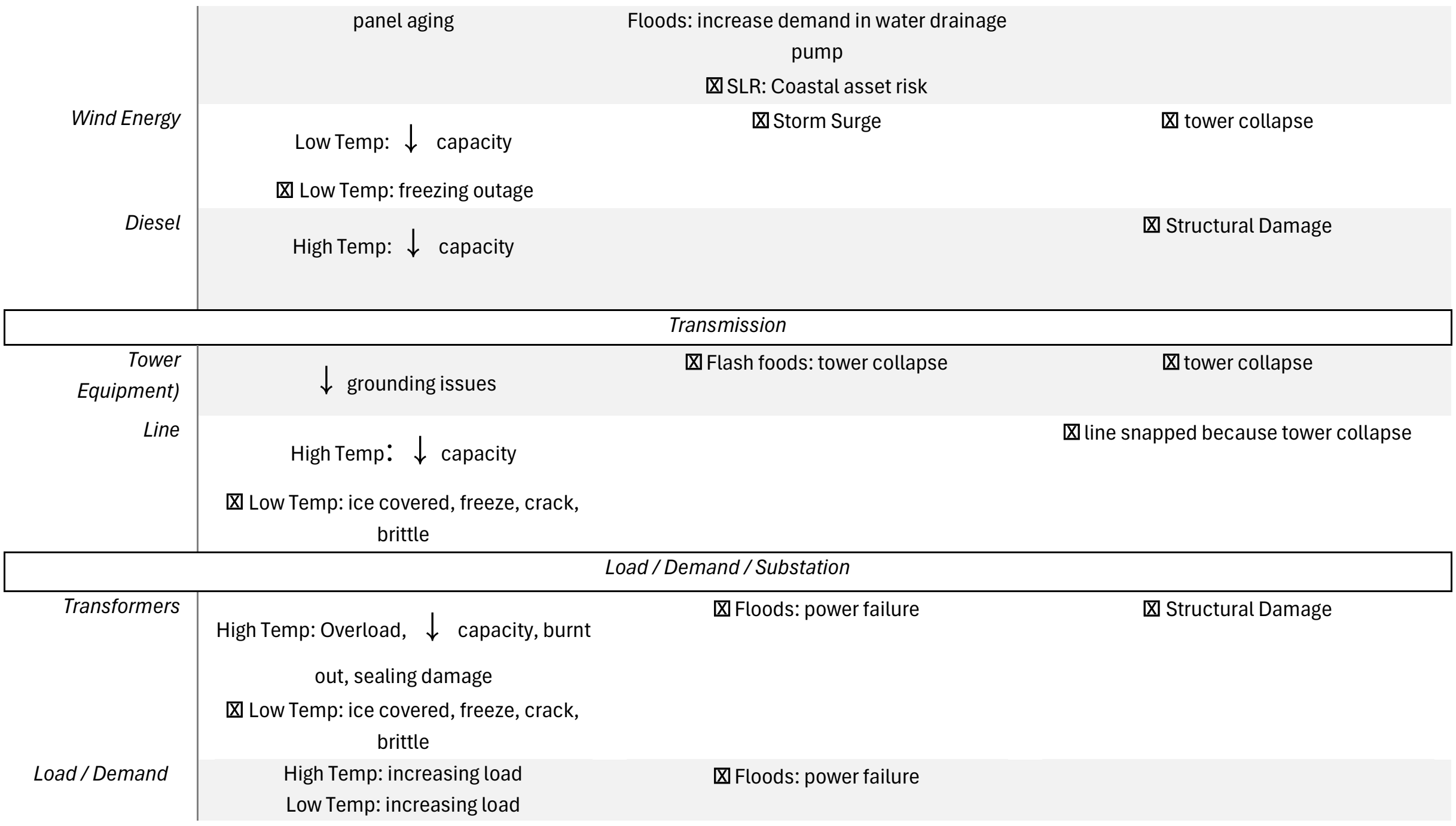

| | | | |
|---|---|---|---|
| | panel aging | Floods: increase demand in water drainage pump<br>☒ SLR: Coastal asset risk | |
| *Wind Energy* | Low Temp: ↓ capacity<br>☒ Low Temp: freezing outage | ☒ Storm Surge | ☒ tower collapse |
| *Diesel* | High Temp: ↓ capacity | | ☒ Structural Damage |
| | | *Transmission* | |
| *Tower Equipment)* | ↓ grounding issues | ☒ Flash foods: tower collapse | ☒ tower collapse |
| *Line* | High Temp: ↓ capacity<br>☒ Low Temp: ice covered, freeze, crack, brittle | | ☒ line snapped because tower collapse |
| | | *Load / Demand / Substation* | |
| *Transformers* | High Temp: Overload, ↓ capacity, burnt out, sealing damage<br>☒ Low Temp: ice covered, freeze, crack, brittle | ☒ Floods: power failure | ☒ Structural Damage |
| *Load / Demand* | High Temp: increasing load<br>Low Temp: increasing load | ☒ Floods: power failure | |

## B. Stressors: Derating Function

Table B. 1 provides a concise summary of the derating functions and corresponding equations implemented in the analytical framework and code. The simple formulations are chosen to provide a first-order approximation of temperature- and climate-related performance losses across generation technologies. These explicit equations are included to ensure transparency, reproducibility, and clarity in the computational implementation of the stressor assessment.

*Table B. 1. Stressor functions with the equation used in the framework and code.*

| *Elements* | *Function* | *Equation* | |
|---|---|---|---|
| *Steam: Coal Fired Power Plant* | ↓ Capacity 0.34 % every ↑ 1°C [a)] | $P_{derated} = P_{rated}(1 - \alpha_{coal} * \max(0, T_w - T_{ref_{coal}}))$<br>$\alpha = 0.0034, T_{ref_{coal}} = 25$ | (1) |
| *Steam: Nuclear* | ↓ Capacity 0.44 % every ↑ 1°C [b)] | $P_{derated} = P_{rated}(1 - \alpha_{nuc} * \max(0, T_w - T_{ref_{nuc}}))$<br>$\alpha = 0.0044, T_{ref_{nuc}} = 25$ | (2) |
| *Gas: Open Cycle (OCGT)* | ↓ Capacity 0.6 % every ↑ 1°C [c)] [d)] | $P_{derated} = P_{rated}[1 - \alpha_{OCGT} \max(0, T - T_{ref,OCGT})]$<br>$\alpha = 0.006, T_{ref_{nuc}} = 16$ | (3) |
| *Gas: Combine Cycle (CCGT)* | ↓ Capacity 0.45 % every ↑ 1°C [c) d)] | $P_{derated} = P_{rated}[1 - \alpha_{OCGT} \max(0, T - T_{ref,OCGT})]$<br>$\alpha = 0.0045, T_{ref_{nuc}} = 16$ | (4) |
| *Diesel* | *Derated capacity due to elevation and air temperature [e)]* | $P_{derated} = P_{rated} \cdot [1 - (a + b + c)]$<br>$a = \alpha_{amb} \cdot \max(0, T_{amb} - T_{ref})$<br>$b = \alpha_{alt} \cdot \max(0, h - h_{ref})$<br>$c = \alpha_{CAC} \cdot \max(0, T_{CAC} - T_{ref,CAC})$<br>$\alpha_{amb} = 0.005, \alpha_{alt} = 0.0001, \alpha_{CAC} = 0.004$ | (5) |
| *PV* | *Derated capacity due to irradiance, air temperature [d)]* | $P_{derated} = P_{rated} * \frac{irr}{1000}\left(1 - \varepsilon(T - T_{ref})\right)$<br>$irr = 1000,\ T_{ref} = 25, \varepsilon = 0.004$ | (6) |
| | Note:<br>$P_{derated}$= Derated capacity<br>$P_{rated}$ = installed capacity<br>α = derate factor | | |

| | | |
|---|---|---|
| | $T_{ref}$ = reference temperature, the starting temperature for derate<br>irr = irradiation<br>$\varepsilon$ = temperature coefficient<br>a,b,c = diesel reduction factor of ambient (amb), altitude (alt), and charge air cooler (CAC).<br>h = diesel generator altitude | |
| *Transmission Lines* | *Thermal energy balance equation IEEE standard 783™ - 2012* [f] | $$I = \sqrt{\frac{q_c + q_r - q_s}{R(T_c)}}$$ $$q_c = \pi D \cdot 10.45 \left(1 + 10\sqrt{\{v\}}\right)(T_c - T_a)$$ $$q_r = \pi D\, \varepsilon\sigma\, (T_c^4 - T_a^4)$$ $$q_s = \alpha\, S\, \pi D$$ *(7)* |
| | Note:<br>$q_c$ = convective heat loss from the conductor to the surrounding air<br>$q_r$ = radiative heat loss from the conductor surface<br>$q_s$ = solar heat gain absorbed by the conductor<br>$R(T_c)$ = electrical resistance of the conductor at allowable conductor operating temperature $T_c$<br>$T_c$ = Allowable conductor operating temperature = 90<br>$T_a$ = ambient air temperature<br>D = outside diameter of the conductor<br>$\varepsilon$ = emissivity of the conductor surface = 0.9<br>σ = Stefan–Boltzmann constant ($5.67 \times 10^{-8}$ $W \cdot m^{-2} \cdot K^{-4}$)<br>α = solar absorptivity of the conductor surface = 0.5<br>S = solar radiation intensity incident on the conductor = 500<br>V = wind speed at the conductor location | |
| *Demand* | *Correlation: Daily energy (MWh) – Daily t2m (°C) regression, controlled with date, day type, and holidays.* | $$MWh_t = \beta^0 + \beta^1 \cdot temp_{max_t} + \sum\nolimits_{Y} \gamma_Y \cdot 1\{Year_t = y\} + \sum\nolimits_{m} \delta_m \cdot 1\{Month_t = m\} + \sum_{\tau} \theta_\tau \cdot 1\{DayType_t = \tau\} + \sum_{d} \kappa_d \cdot 1\{DayName_t = d\} + \eta^1 \cdot Weekend_t + \eta^2 \cdot Holiday_t + \eta^3 \cdot Ext_{Holiday_t} + \varepsilon_t$$ *(8)* |
| | *Future projection, based on annual temperature comparison, demand group, and correlation* | $$D_{adj(p,t)} = D_{base(p,t)} \cdot \left[1 + \beta_p \left(\frac{RCR_{(p,t)}}{RCR_{(p,2024)}}\right) \Delta T_{(p,t)}\right]$$ |
| | Note:<br>$MWh_t$ = electricity demand (MWh) at time t<br>$\beta^0$, $\beta^1$ = regression intercept and temperature coefficient<br>$\gamma_y$ = year fixed-effect coefficient<br>$\delta_m$ = month fixed-effect coefficient<br>$\theta_t$ = day-type fixed-effect coefficient<br>$\kappa_d$ = day-of-week fixed-effect coefficient<br>$\eta^{1,2,3}$ = coefficients for weekend, holiday, and extended-holiday effects<br>$\varepsilon_t$ = regression error term | |
| | $D_{adj(p,t)}$ = temperature-adjusted electricity demand in province p and year t (GWh)<br>$D_{base(p,t)}$ = baseline electricity demand projection without climate adjustment (GWh)<br>$\beta_p$ = demand–temperature sensitivity coefficient for province p,<br>$RCR_{(p,t)}$=Ratio of Commercial and Residential demand group in province p and year t<br>$\Delta T_{(p,t)}$ = change in ambient temperature between the future climate projection (e.g., RCP scenario) and the historical baseline derived from ERA5 | |

*a) (Petrakopoulou et al., 2020)*
*b) (Attia, 2015)*
*c) (Handayani et al., 2020)*
*d) (Bartos & Chester, 2015)*
*e) (Wärtsilä Diesel Oy, n.d.)*
*f) (IEEE, 2013)*

## B.1. Stressors: Generation Derate

Steam-based power plants located in coastal areas typically rely on seawater for cooling, meaning that condenser performance is influenced by water temperature. To examine whether ambient air temperature can reasonably represent cooling conditions, we compare hourly near-surface air temperature (t2m) and sea surface temperature (SST) from ERA5 (2024) at several representative coastal generation sites as depicted in Figure B. 1. The results show clear temporal similarity between the two variables, including comparable seasonal patterns and coinciding peak periods. Although SST varies more smoothly due to ocean thermal inertia, its overall range and timing closely follow ambient air temperature.

Based on this observed similarity and to maintain consistency with other generation technologies—whose derating functions are defined using ambient temperature—we adopt near-surface air temperature (t2m) as the climatic input for steam turbine derating calculations in both historical and future scenarios. This choice ensures methodological consistency across technologies while remaining physically plausible for national-scale assessment, where the objective is to capture general thermal stress rather than site-specific condenser dynamics.

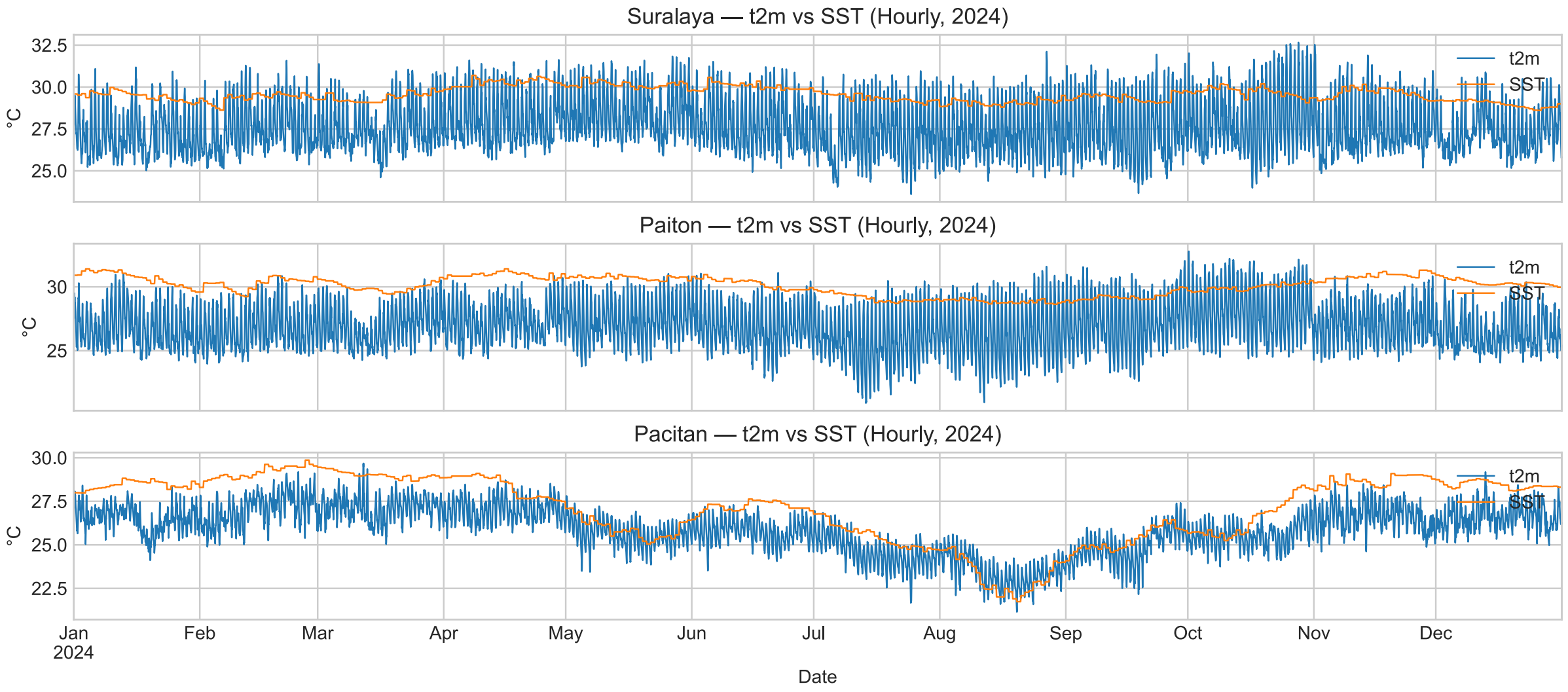

*Figure B. 1 Daily maximum air temperature and sea surface temperature from ERA5 in year 2024.*

To assess the sensitivity of the derating results under the long-term plan, RUKN, in year 2060, we conduct a temperature sensitivity analysis by varying the assumed national average temperature, depicted in Figure B. 2. Instead of relying on a single projected value, this experiment explores a range of plausible temperature conditions to evaluate how generation capacity losses respond to incremental thermal stress. The experiment uses the Exploratory Modeling Workbench (Kwakkel, 2017), available on GitHub.

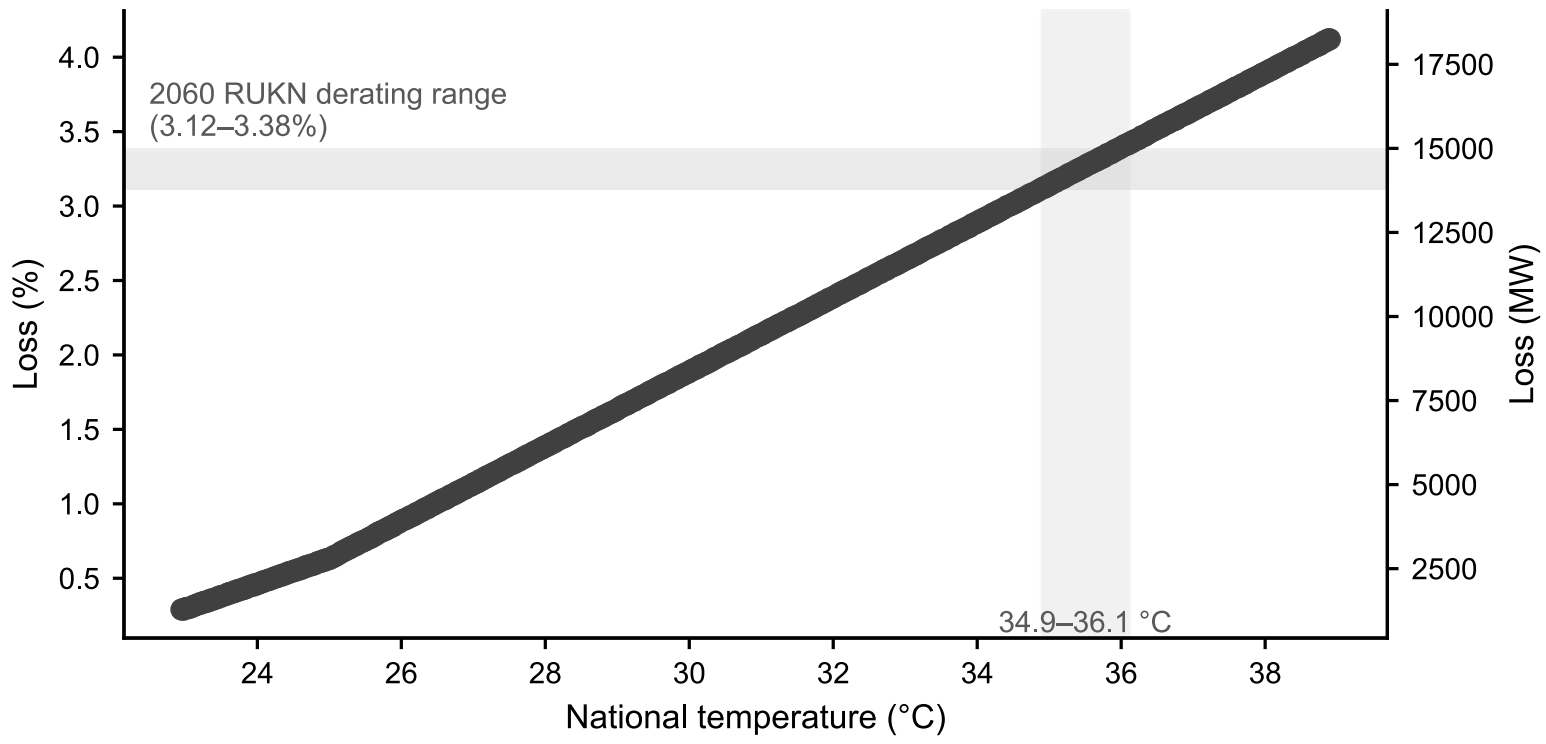

*Figure B. 2 Sensitivity of generation capacity losses in the 2060 RUKN plan to variations in assumed national average temperature.*

We examined the long-term plan, RUKN, with the maximum tasmax retrieved for the corresponding year from the CMIP5 climate model. As shown in the right box of Figure B. 3, under these conditions, temperature-related derating is estimated at 2.69–2.83% from total capacity, resulting in losses of 11.9–12.5 GW for RCP 4.5 and RCP 8.5, respectively. RCP 8.5 projected warmer temperatures, thus the impact on the generator derating is higher.

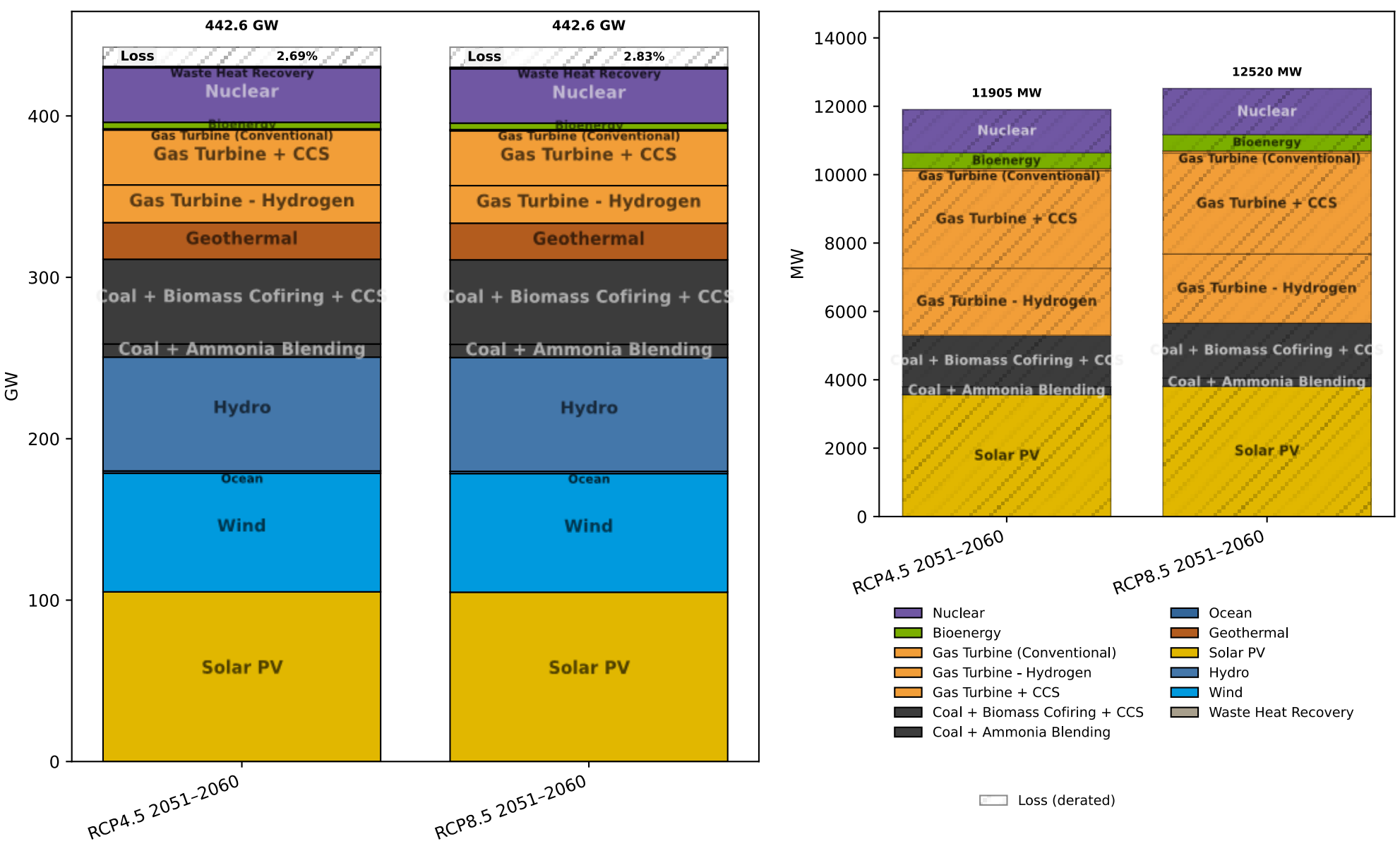

*Figure B. 3 The long-term plan, RUKN, in the year 2060. The left subplot shows the total planned installed capacity, and the right subplot shows the temperature-based loss portion.*

For the 10-year plan, as delivered in Figure B. 4, the projected derating reaches 3.35-3.42% under REBase in the RCP 4.5 and 8.5 scenarios, respectively. Under ARED, the derate is projected to be 3.10-3.17% in the RCP4.5 and 8.5 scenarios. These percentages are comparable to the previous 2060 derate. They appear small, yet correspond to losses of 4.3-4.6 GW—an impact that is far from negligible.

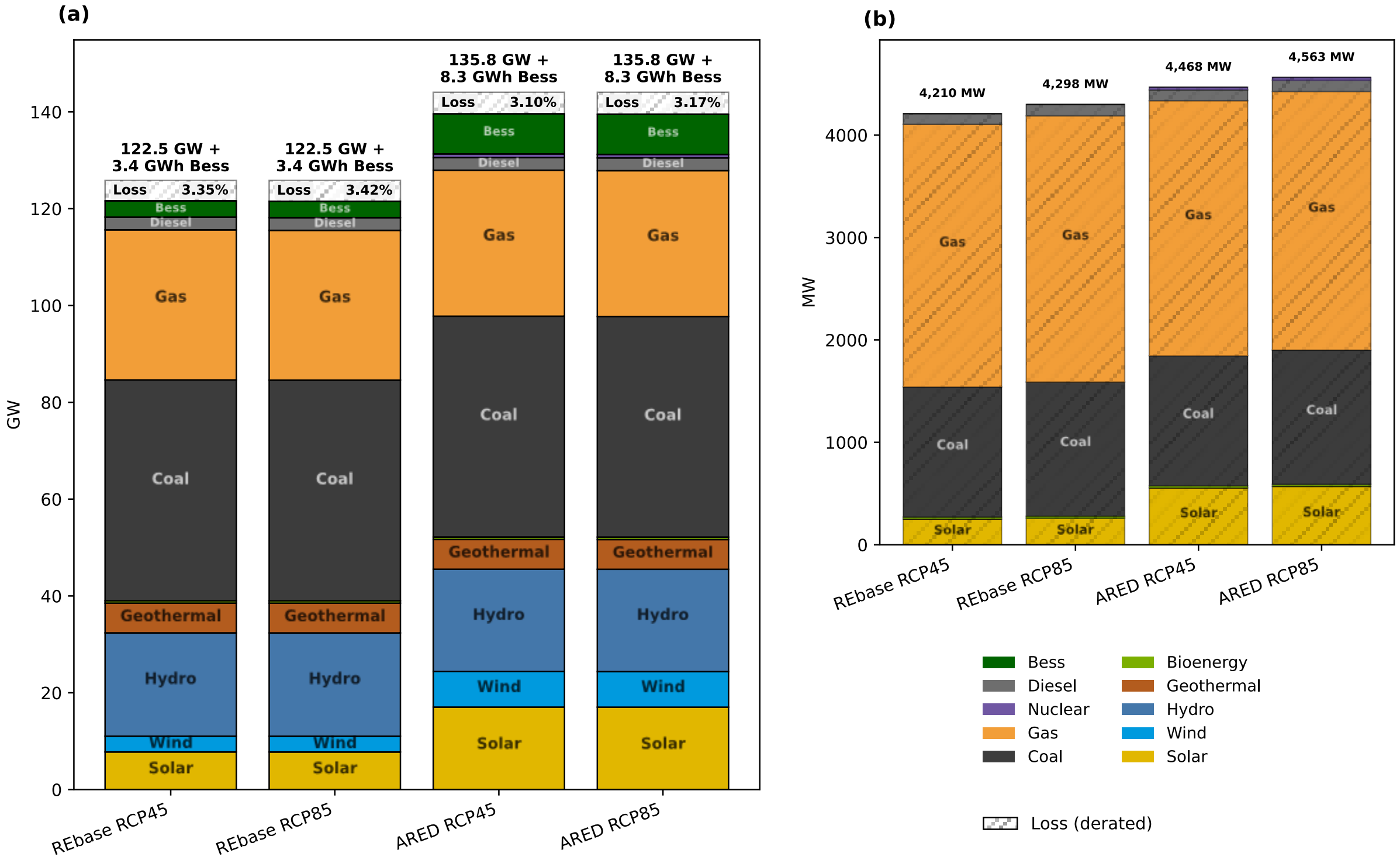

*Figure B. 4 The 10-year plan assessment in the year 2034. Subplot a shows the total planned installed capacity by 2034, and subplot b shows the loss portion due to temperature.*

## B.2 Stressors: Transmission Lines Derate

To parameterize wind-related transmission line derating, we derive representative wind speed conditions from spatial data. Figure B. 5 presents the distribution of 10 m wind speed from ERA5 in year 2024, extracted along all transmission line segments across Indonesia. Wind values are obtained through spatial overlay between the ERA5 raster grid and the national transmission network dataset, resulting in approximately 1.6 million samples.

The empirical distribution is used to define low- and high-wind reference conditions. Specifically, the 10th percentile (P10) represents low-wind conditions, while the 90th percentile (P90) represents high-wind conditions. These percentile thresholds provide standardized boundary conditions for transmission ampacity calculations and allow consistent comparison across future analysis. The approach generalizes wind exposure at the national scale rather than modeling site-specific extreme events.

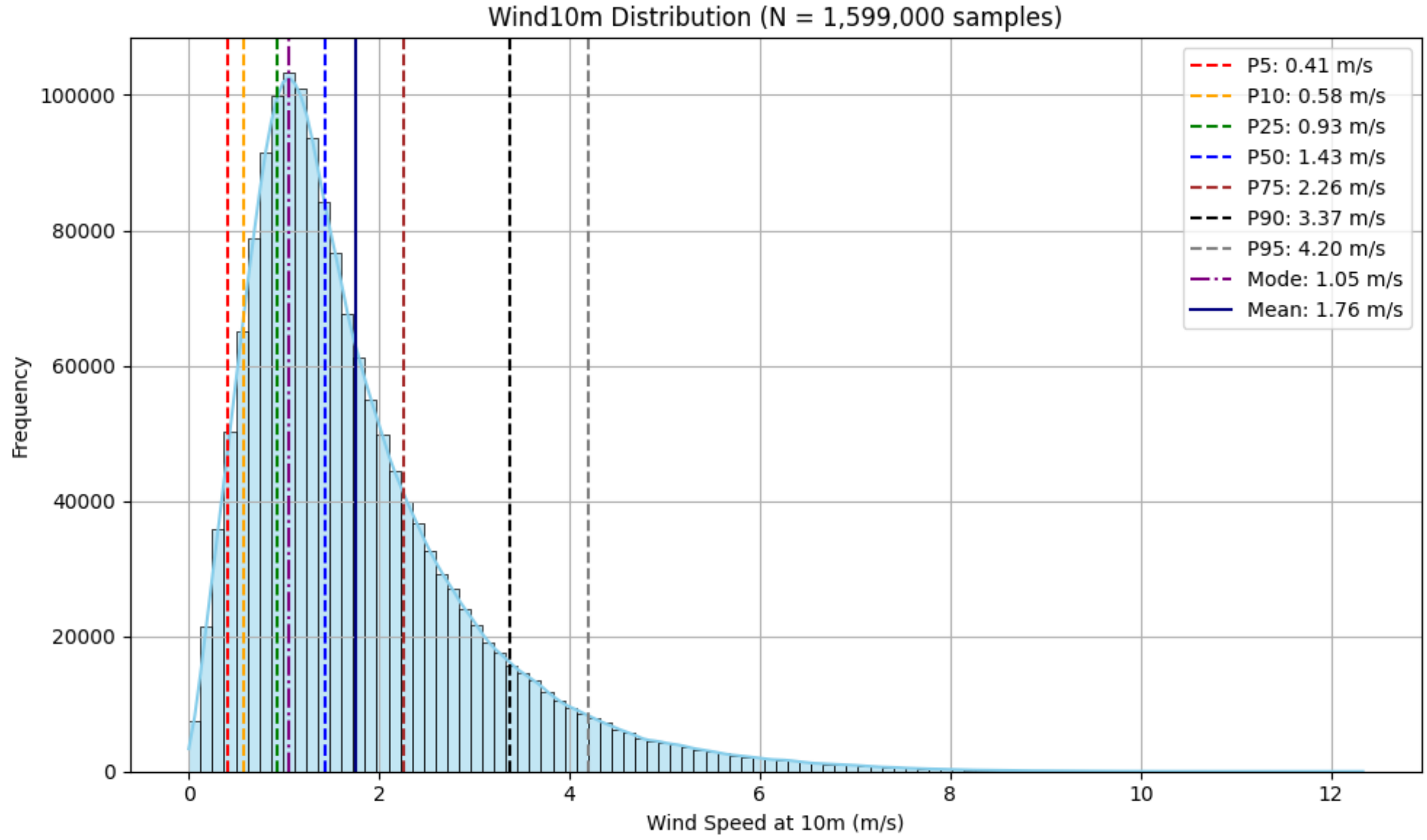

*Figure B. 5 The wind speed from ERA5 2024 across every generation asset in Indonesia. We sample the distribution of wind speed to generalize the statistics of wind speed. We use low wind conditions of P10 and high wind conditions of P90.*

## B.3 Stressors: Demand Change

The regression analysis between temperature and electricity demand is based on daily system-level load data obtained from PLN and near-surface air temperature from ERA5. For each regional system, daily electricity demand is paired with the corresponding daily temperature values to examine the empirical relationship between thermal conditions and system load. Owing to data-sharing limitations, the raw dataset cannot be disclosed. Therefore, only the plot is presented in Figure B. 6. The figure presents the daily time series for selected systems, allowing visual inspection of co-movements and overall correlation patterns between temperature and demand. The coefficient of determination ($R^2$) is reported for each system to indicate the strength of the temperature–demand relationship used in the regression model.

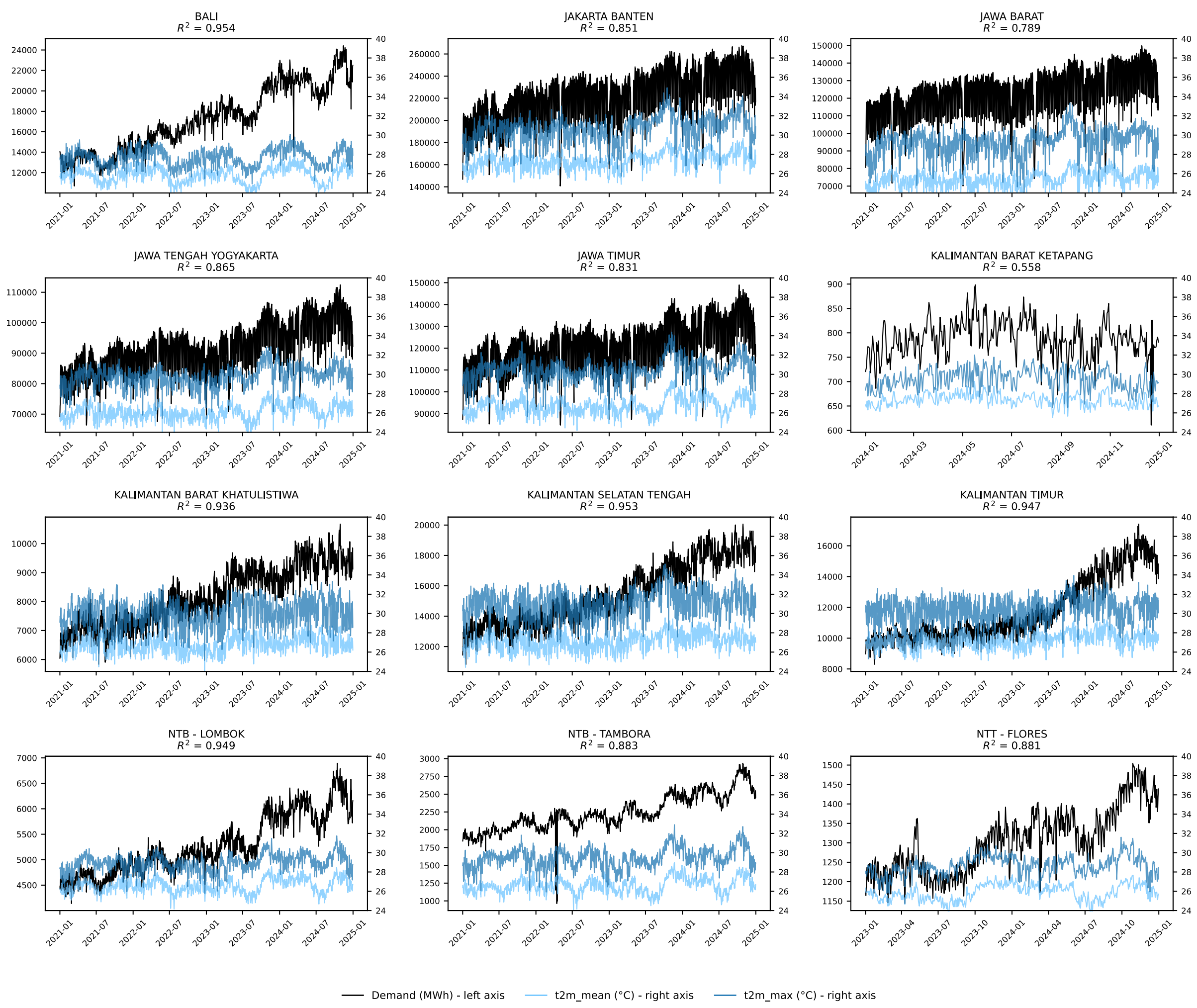
BALI
$R^2$ = 0.954
JAKARTA BANTEN
$R^2$ = 0.851
JAWA BARAT
$R^2$ = 0.789
JAWA TENGAH YOGYAKARTA
$R^2$ = 0.865
JAWA TIMUR
$R^2$ = 0.831
KALIMANTAN BARAT KETAPANG
$R^2$ = 0.558
KALIMANTAN BARAT KHATULISTIWA
$R^2$ = 0.936
KALIMANTAN SELATAN TENGAH
$R^2$ = 0.953
KALIMANTAN TIMUR
$R^2$ = 0.947
NTB - LOMBOK
$R^2$ = 0.949
NTB - TAMBORA
$R^2$ = 0.883
NTT - FLORES
$R^2$ = 0.881
Demand (MWh) - left axis
t2m_mean (°C) - right axis
t2m_max (°C) - right axis

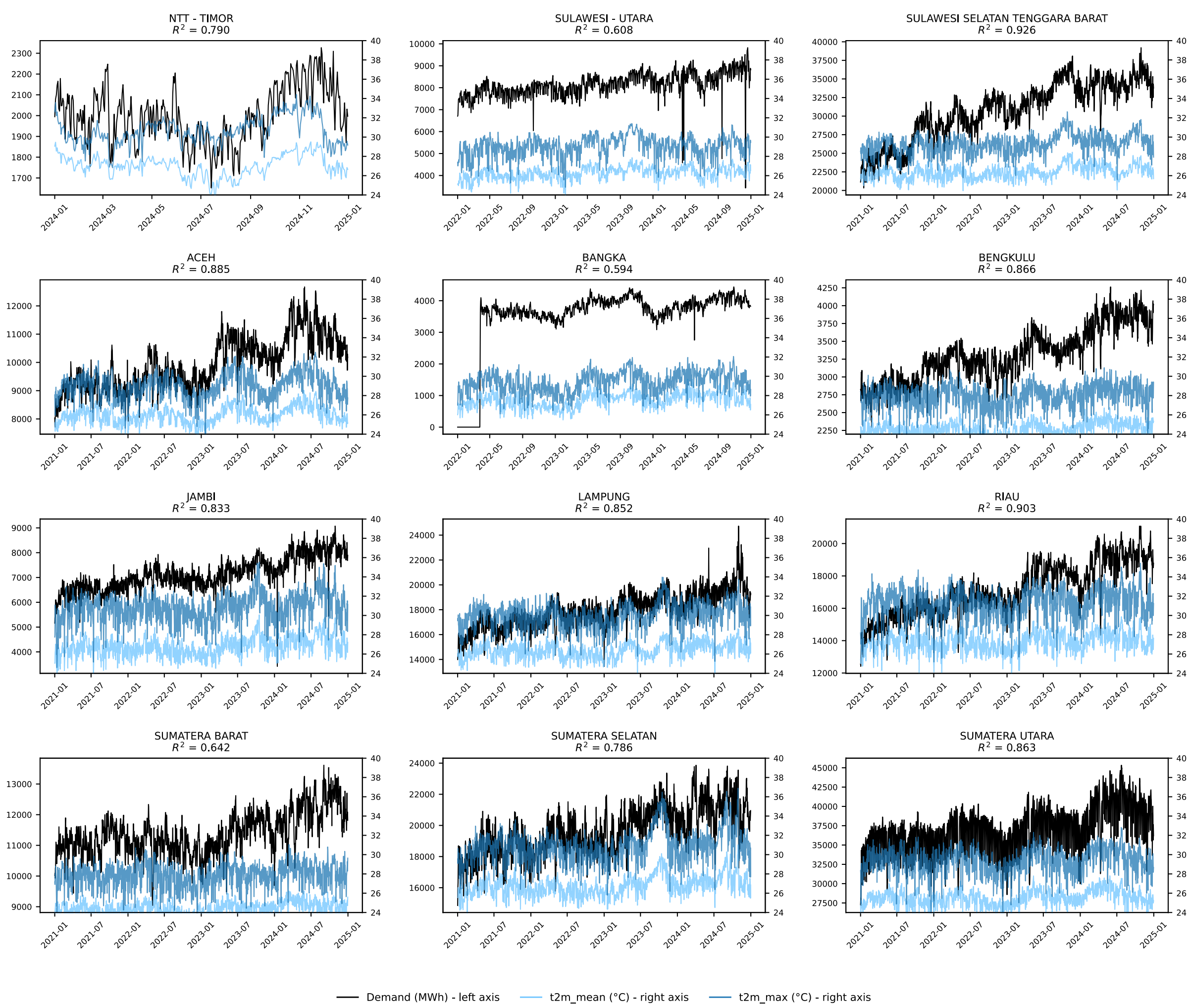

*Figure B. 6 Daily time series of regional electricity demand, along with mean and maximum daily temperatures from ERA5.*

## C. Shocks: Summary of Floods, Landslides, and Sea Level Rise.

Figure C. 1 provides the detailed numerical hazard index values underlying the categorical results presented in the main paper. While the main text groups exposure levels into qualitative categories for clarity, this figure reports the full index values for floods, landslides, and long-term sea-level rise across provinces and asset types. The heatmaps, therefore, allow transparent inspection of the magnitude of exposure and the relative differences between regions.

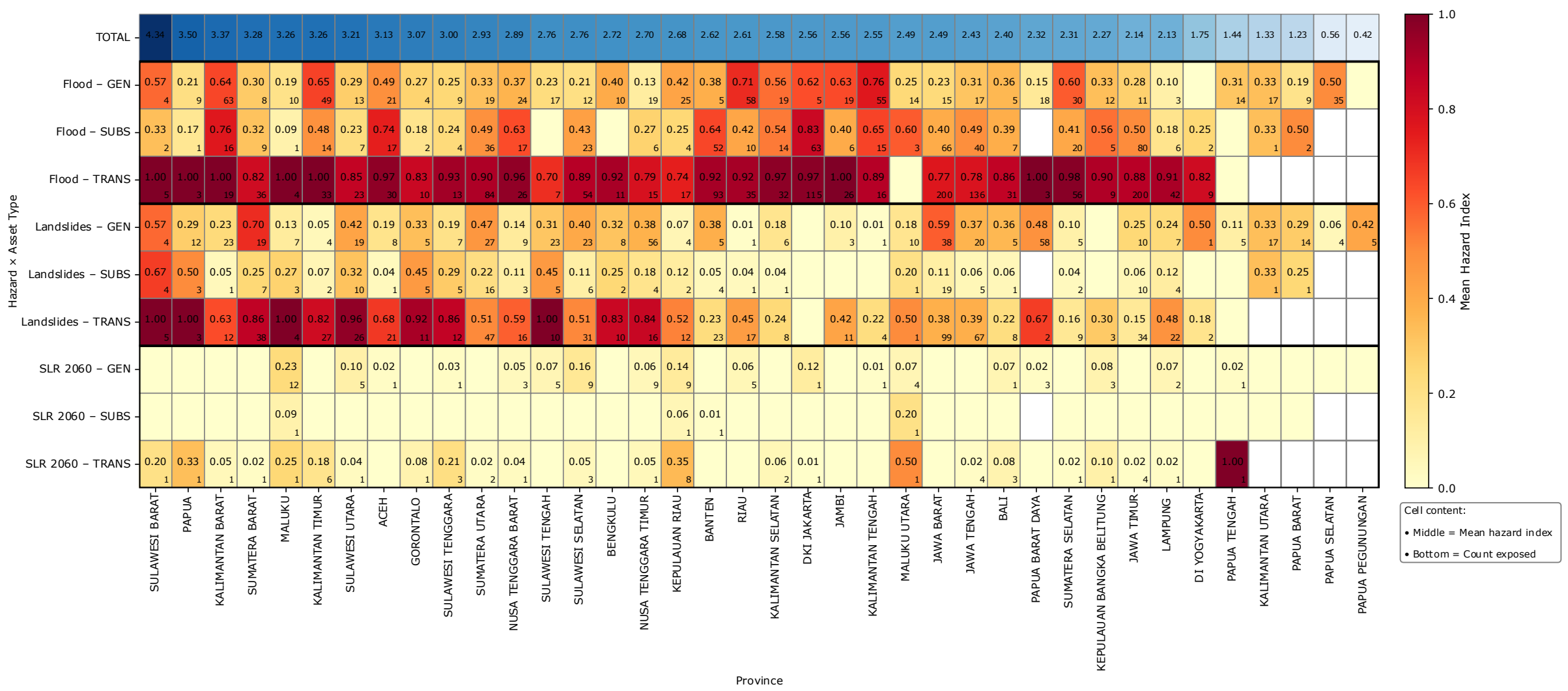

*Figure C. 1 Hazard heatmaps, asset exposure index, and counts from floods, landslides, and long-term sea level rise. The top row summarizes the total hazard index per province. For floods and landslides, exposure is based on 2024 InaRISK hazard indices and assets, while sea level rise is based on 2060 projections applied to 2024 assets. Heatmaps indicate the number of exposed assets. If no index is shown or appears, the hazard index is equal to 0.00. Pure white areas indicate no assets in the GIS dataset.*

To assess the long-term extension of sea level rise to 2150 using projections from the IPCC AR6 dataset, Figure C. 2 illustrates the cumulative inundation of power system assets under alternative SSP pathways and associated uncertainty bounds.

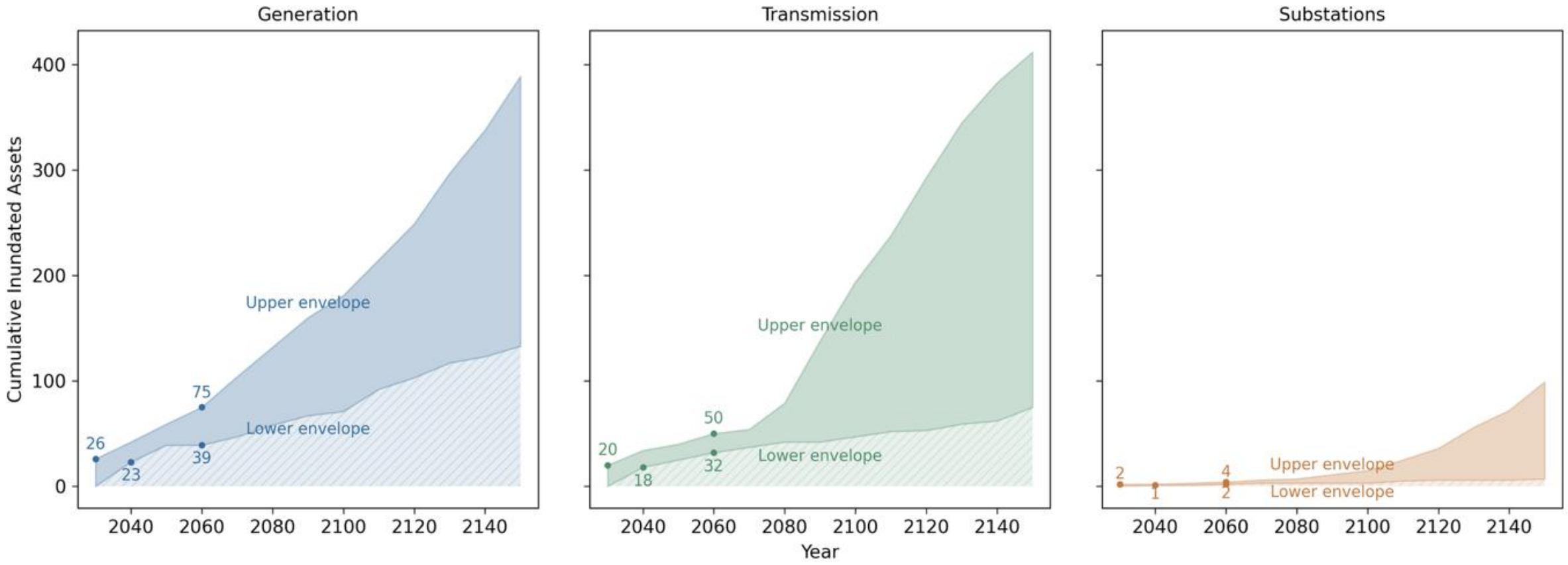

*Figure C. 2 National cumulative exposure of electricity infrastructure to sea-level rise (SLR) up to the year 2150 using the IPCC AR6 dataset. The upper envelope represents the highest projected count of inundated assets across SSP2-4.5, SSP3-7.0, and SSP5-8.5 under the high-end estimate (95th percentile), while the lower envelope represents the corresponding median projections (50th percentile) across the same scenarios.*

## D. Compound Stressor-Shock Provincial Analysis.

While the main paper focuses on reserve margins at the major interconnected system level, Figure D 1 provides a more detailed provincial comparison of electricity supply and demand under compound stressor–shock conditions in the 10-year plan scenario in year 2034. It contrasts generation capacity after temperature- and hazard-induced losses with the original planned capacity and temperature-driven demand increases, highlighting how climate impacts tighten provincial supply–demand balances and propagate to system-level reserve margins across major interconnected systems. Then, Figure D. 1 visualizes the provincial reserve margin in a stacked bar view to give the overall impression of each province.

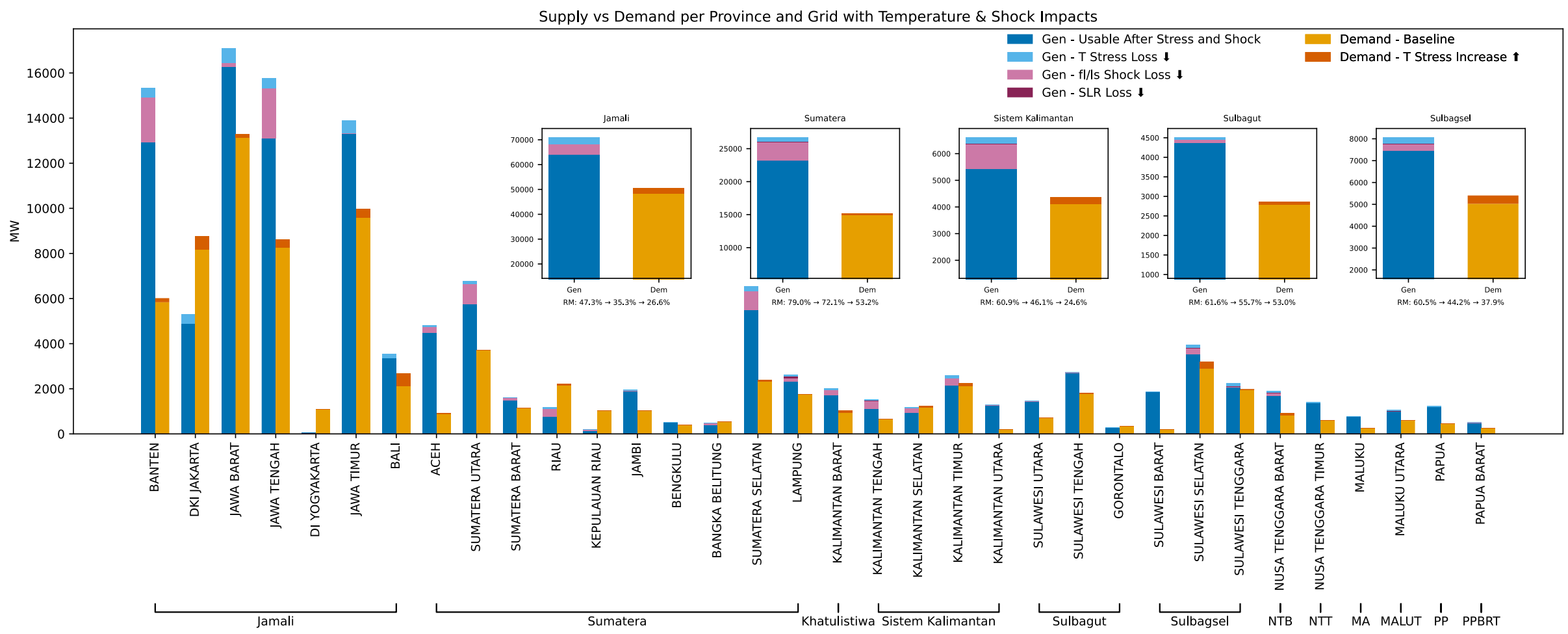

*Figure D. 1. Detailed provincial comparison of supply and demand under compound stressor–shock impacts in the 10-year plan, RUPTL 2034 REBase scenario, showing generation capacity after temperature and hazard losses alongside baseline and temperature-driven demand increases. Insets summarize the resulting reserve margins for major interconnected systems.*

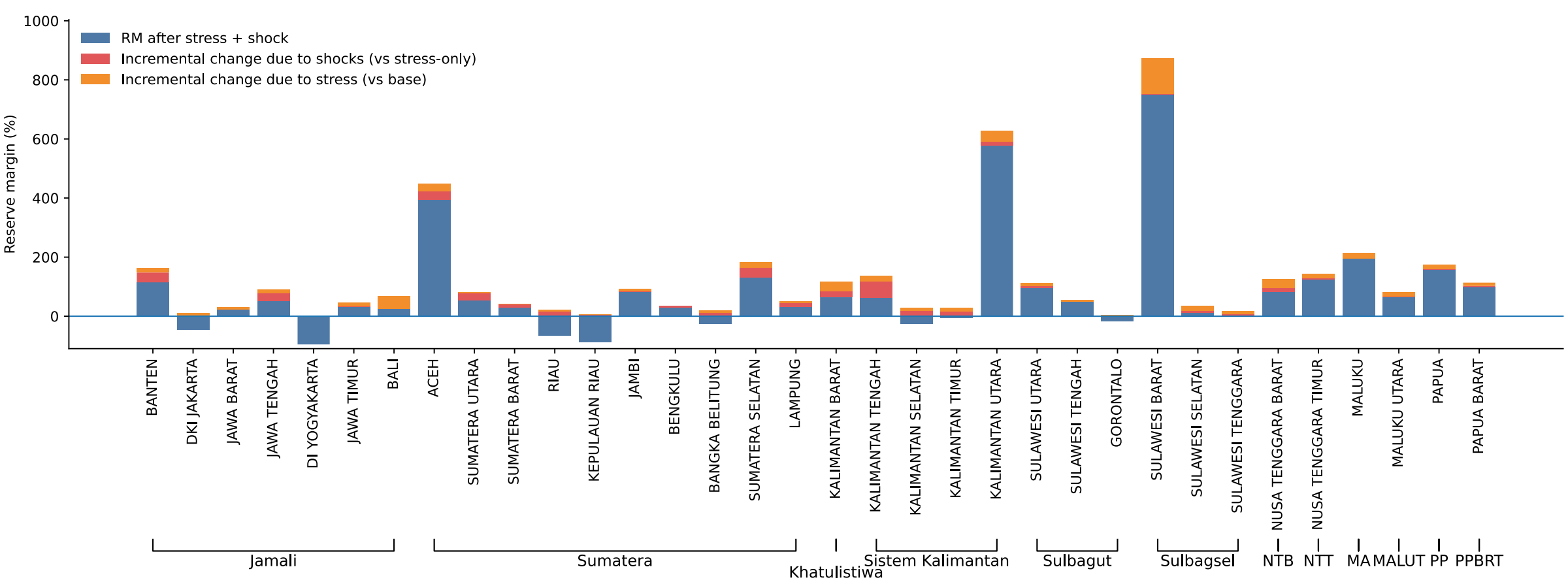

*Figure D. 1 Stacked view of provincial reserve margin under compound stressors-shocks.*